\documentclass[reprint,amsmath,amssymb,aps]{revtex4-2}

\usepackage{graphicx}
\usepackage{bm}
\usepackage[hidelinks]{hyperref}
\usepackage{amsthm}
\usepackage{color}

\theoremstyle{plain}

\theoremstyle{plain}

\providecommand{\corollaryname}{Corollary}
\providecommand{\theoremname}{Theorem}

\newcommand{\Tr}{\mathrm{Tr}}

\begin{document}
\title{Coherent Fluctuations in Noisy Mesoscopic Systems, the Open Quantum SSEP and Free Probability}

\author{Ludwig Hruza}
\email{ludwig.hruza@ens.fr}
\author{Denis Bernard}%
 \email{denis.bernard@ens.fr}

\affiliation{\vskip 0.5 truecm Laboratoire de Physique de l'\'Ecole Normale Sup\'erieure, CNRS, ENS \& PSL University, Sorbonne Universit\'e, Universit\'e Paris Cit\'e, 75005 Paris, France}%

\date{\today; First submission: May 4, 2022}

\begin{abstract}
Quantum coherences characterise the ability of particles to quantum mechanically interfere within some given distances. In the context of noisy many-body quantum systems these coherences can fluctuate. A simple toy model to study such fluctuations in an out-of-equilibrium setting is the open quantum symmetric simple exclusion process (Q-SSEP) which describes spinless fermions in one dimension hopping to neighbouring sites with random amplitudes coupled between two reservoirs. Here we show that the dynamics of fluctuations of coherences in Q-SSEP have a natural interpretation as free cumulants, a concept from free probability theory. Based on this insight we provide heuristic arguments why we expect free probability theory to be an appropriate framework to describe coherent fluctuations in generic mesoscopic systems where the noise emerges from a coarse-grained description. In the case of Q-SSEP we show how the link to free probability theory can be used to derive the time evolution of connected fluctuations of coherences as well as a simple steady state solution.
\end{abstract}

\maketitle

\tableofcontents{}

\section{Introduction}

From the point of view of statistical physics, a system is completely characterised if the probability of all possible microscopic configurations is known. From this knowledge one can obtain the probability distribution of macroscopic (or thermodynamic) variables such as the density or the current profile, which are in principle experimentally observable. In an equilibrium situation, where the probability distribution on microscopic configurations is well known since Boltzmann, macroscopic variables satisfy a large deviation principle: they are strongly peaked at their mean value around which their probability distribution decays exponentially with the number of particles in the system and proportional to the appropriate thermodynamic potential (e.g. the entropy or the free energy), see e.g. \cite{Touchette2009}.

In contrast to this, out-of-equilibrium situations might depend on a large variety of system dependent details and there are no general formulas for the probability distribution on microscopic configurations. Nevertheless, over the course of the last 30 years a lot of progress has been made to understand the statistics of macroscopic variables in out-of-equilibrium systems, which show rich and new features. For example, density correlations in non-equilibrium steady states extend to macroscopic distances, as has been experimentally observed by Dorfmann \cite{Dorfman1994Generic}. Furthermore, fluctuations relations that go beyond the linear response regime, have been shown to be generically applicable to out-of-equilibrium systems. \cite{Jarzynski1997Nonequilibrium,Crooks1999Entropy}. And the large deviation principle introduced above in the equilibrium context has been realised to hold also to the non-equilibrium setting -- with the need to formulate appropriate out-of-equilibrium thermodynamic potentials, see e.g. \cite{Derrida2007Non-equilibrium}. 

For classical systems these efforts have culminated in the formulation of the so-called \textit{macroscopic fluctuation theory} (MFT) \cite{Bertini2005Current,Bertini2015MFT} which applies to systems with diffusive transport. Usually the system is maintained out-of-equilibrium by coupling the boundaries to reservoirs at different chemical potentials. MFT allows to specify the probability distribution of the density and current profile in such systems and remarkably, this relies on only two system-dependent quantities, the diffusion constant and the mobility. The development of MFT has been strongly inspired by the study of microscopic toy models, more precisely, stochastic lattice gases, that revealed universal properties in the density and current fluctuations in the sense that they did not depend on the precise underlying microscopic dynamics. A very important role in this context has been played by the symmetric and asymmetric simple exclusion processes (SSEP and ASEP), since these toy models are exactly solvable \cite{Derrida1993Exact,Derrida1998An,Derrida1999Bethe,Bodineau2004Current} (for a review, see \cite{Derrida2007Non-equilibrium,Mallick2015TheExclusion}).

The major question we are concerned with is whether MFT can be extended to a quantum setting \cite{Bernard2021CanMFT}. Such a theory, which could be called the {\it quantum mesoscopic fluctuation theory}, should not only describe the statistical properties of the diffusive transport (i.e. density and current profiles) in out-of-equilibrium quantum systems, but also those of quantum coherent effects such as interference or entanglement. These effects are inscribed into the \textit{coherences} $G_{ij}=\Tr(\rho c_i^\dagger c_j)$, the off-diagonal elements of the density matrix $\rho$ when expressed in the particle number basis on each site $i,j=1,\cdots,L$. A big difference to classical noisy systems (such as SSEP) is that the density matrix in a noisy quantum system -- usually interpreted as the probability distribution of quantum states -- is itself a dynamical variable that fluctuates. A quantity sensible to these fluctuations is the entanglement entropy since it is composed from partial traces of powers of $\rho$. Fluctuations of the entanglement entropy have already been studied in the context of unitary random circuits \cite{Nahum2017Quantum,Nahum2018Operator} (see \cite{Fisher2022Random,potter2021entanglement} for a review), as well as certain toy models of driven diffusive systems \cite{Gullans2019Entanglement}. In seeking a mesoscopic fluctuation theory we hope to establish a general framework to deal with coherent effects in diffusive systems, based on coherences $G_{ij}$.

\begin{figure}\label{fig:picture}
\begin{center}
\includegraphics{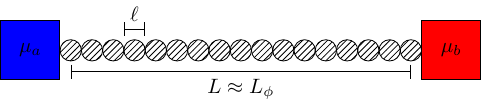}
\caption{ A schematic representation of a mesoscopic system coupled to two reservoirs of different chemical potentials $\mu_a$ and $\mu_b$. Here $\ell$ denotes the ballistic length, above which the transport is diffusive, and $L_\phi$ the coherence length, below which interference effects can be observed.}
\end{center}
\end{figure}

The picture (see Fig.\ \ref{fig:picture}) we have in mind is that such a quantum extension of MFT should apply to \textit{mesoscopic systems}, i.e.\ to systems in which the system size $L$ is of the order of the coherence length $L_\phi$ of the dynamical degrees of freedom (e.g. for electrons in a disordered metal of size $L_\phi\approx 1\mathrm{\mu m}$ \cite{Steinbach1996Observation}). Another important length scale is the mean free path $\ell$, below which transport is ballistic and the microscopic degrees of freedom change very rapidly with time. After an average over the microscopic degrees of freedom inside \textit{ballistic cells} of the size of the mean free paths $\ell$, we expect that the system has a coarse-grained description in terms of a stochastic and unitary process. These ballistic cells should be assumed to be thermodynamically large, but small compared to the system size, $1\ll\ell\ll L$ (we identified the physical length with the number of lattice sites by setting the lattice constant to one, $a_\mathrm{uv}=1$). At length above $\ell$ we expect diffusive transport, and at length below $L_\phi$ we observe quantum mechanical interference. This is the mesoscopic regime we are interested in.

The decomposition of a many-body quantum systems into thermodynamically large cells finds many examples in the literature. Notably, it appears as the main feature of Generalized Hydrodynamics (GHD) which has allowed to study out-of-equilibrium dynamics in integrable systems \cite{Bertini2016Transport,Castro2016Emergent} (see  \cite{Doyon2020Lecture,Alba2021Generalized} for a review). Since this theory treats the system as a composition of many thermodynamically large fluid cells that are locally in equilibrium with respect to a generalised Gibbs ensemble, it looses the statistical properties of quantum correlations between the fluid cells. Note that an approach to restore these quantum correlations in GHD has recently been proposed in \cite{Ruggiero2020Quantum}. Furthermore, GHD is mostly concerned with ballistic transport due to the integrable nature of the systems to which it applies. It therefore does not seem to be a good candidate for a quantum version of MFT.

A toy model that allows to study fluctuations in the mesoscopic regime is the quantum symmetric simple exclusion process (Q-SSEP) \cite{Bauer2017Stochastic,Bauer2019Equilibrium,Bernard2019Open,Bernard2021Solution,Bernard2021Entanglement,Bernard2022DynamicsClosed}. It describes the noisy and coherent hopping of spinless fermions on a one dimensional and discrete lattice chain which is maintained out-of-equilibrium by two particle reservoirs at different chemical potentials. While a simple average over the noise completely destroys coherent effects and reduces the model to the classical SSEP, the fluctuations of coherent effects survive in the non-equilibrium steady state \cite{Bernard2019Open}, although they are sub-leading in the system size. It is in this sense that Q-SSEP describes mesoscopic transport. Furthermore, fluctuations of coherences satisfy a large deviation principle in analogy to macroscopic variables in MFT. This can be seen as a first evidence that the fluctuations in Q-SSEP show universal properties that might apply to a larger class of mesoscopic systems. From this point of view a single lattice site in Q-SSEP would correspond to a ballistic cell of the coarse-grained mesoscopic system.

In this paper we show that the mathematical structure of Q-SSEP can be described within the framework of \textit{free probability theory}, an extension of classical probability theory to non-commutative random variables pioneered by Dan Voiculescu in the 90's \cite{Voiculescu1997Free}. More precisely, connected fluctuations of coherences in Q-SSEP correspond to free cumulants which generalise the notion of cumulants from classical probability. Free cumulants have a combinatorial nature and can be obtained from the moments of a random variable as a sums over non-crossing partitions. This is a structure of which we make repeated use in this paper: Firstly, it allows us to derive the time evolution of connected fluctuations in Q-SSEP in a diffusive scaling limit (for the closed Q-SSEP, i.e.\ without boundaries, this scaling limit has been discussed in \cite{Bernard2022DynamicsClosed}). And secondly, it also allows to find a simple derivation for the steady state solution of the fluctuations of coherences. The connection between free probability and the steady state solution of Q-SSEP has already been observed by the mathematician Philippe Biane \cite{Biane2021Combinatorics}. Here we generalise his result to all times $0<t<\infty$ and give a more physical explanation for why free probability arises in our context. In particular, we are able to extend this argument to generic mesoscopic systems which implies that free probability could be the natural mathematical framework to characterise fluctuations in such systems.

This is the main point of the paper: Tools from free probability might play a significant role in understanding fluctuating many-body quantum systems in the mesoscopic regime, both in and out-of-equilibrium. 

Indeed, free probability theory finds an explicit realisation of its concepts through large random matrices \cite{Voiculescu1991Limit,Guionnet2009ICMP,Mingo2017Free}. For a generic mesoscopic system, the coarse-grained view of Fig.\ \ref{fig:picture} in terms of ballistic cells, which undergo a very rapid (unitary) evolution -- much faster than the system's evolution as a whole, can be modelled by large random matrices whose size scales with $\ell$ and whose distribution is invariant under the rapid evolution of the ballistic cell, i.e.\ by the some unitary group (for the moment we leave open the question which degrees of freedom inside the ballistic cells are modelled precisely). In so-far it is not surprising that fluctuations on mesoscopic scales, i.e.\ between ballistic cells, show signs of free probability. We believe that its origin simply lies in a coarse-grained description of microscopic spatial scales and unitary invariance at these scales. Further evidence for this interpretation is provided by the recent observation of free probability in the context of ETH, where instead of spatial scales the coarse-graining is over small energy scales \cite{Pappalardi2022ETH} (submitted simultaneously with our work). 

The structure of the paper is as follows. Section \ref{sec:general_picture} provides heuristic arguments that show why the fluctuations of spatial coherences in generic mesoscopic systems could be appropriately described within free probability theory. In section \ref{sec:free_probability} we give a small introduction to the subject of free probability and in particular explain the relevance of crossing and non-crossing partitions. Section \ref{sec:Q-SSEP} deals with the toy model Q-SSEP. In \ref{subsec:intro_Q-SSEP} we recall the main properties of Q-SSEP. Sections \ref{subsec:Bulk-vs-Boundary} and \ref{subsec:open_boundary_scaling_limit} are concerned with the formulation of the dynamical equations of the correlation functions of the open Q-SSEP in the diffusive scaling limit. Finally, sections \ref{subsec:free_cumulants_in_qssep} explains how to identify free cumulants within Q-SSEP. And \ref{subsec:steady_state_solution} explain how the relation to free cumulants can be used to find a simple steady state solution. Details and proofs are given in a few appendices. In particular, \ref{app:non-crossing_partitions} and \ref{app:derivation-connected-correlations} constitute the major derivations of the free probability structure in Q-SSEP.

\section{General picture}\label{sec:general_picture}
We expect that there is a very general relationship between coherent fluctuations in diffusive one-dimensional fermionic systems in the mesoscopic regime and the mathematical framework of free probability. The aim of this section is to convey an intuition for this claim, not to give complete proofs. We stress that the actual chronological order of our work is the opposite. First it was observed that the mathematical structure of the toy model Q-SSEP has a relation to free probability. Then we realised that one can reduce this relation to three simple conditions which apply to generic mesoscopic systems.

Before stating them some explanation is necessary. In the spirit of Fig.\ \ref{fig:picture} we need to assume that the system exhibits a separation of time scales responsible for fast ballistic transport on scales of the mean free path $\ell$ (inside ballistic cells) versus slow diffusive transport on scales of the system size $L$. In a coarse-grained description of the slow degrees of freedom, the fast ones will act as a source of noise. Mesoscopic observables become therefore random variables with a (yet to be defined) expectation value $\mathbb E_t$. If such observables do not depend on long range coherences, such as the local particle number $\hat n_i:=c_i^\dagger c_i$ ($c_i^\dagger$ is a fermionic creation operator on site $i=1,\cdots,L$) or the associated particle current, then we assume that they are well described by classical MFT, with the replacement $\mathbb E_t[\Tr(\rho\,\bullet)]=\langle \bullet \rangle_t^\text{MFT}$.
In contrast to this, we are interested in the statistical properties of coherences
\[
G_{ij}(t)=\mathrm{Tr}(\rho_{t}c_{i}^{\dagger}c_{j}),
\]
a purely quantum mechanical property without classical analogue -- and therefore outside the scope of classical MFT. In appendix \ref{app:Measuring-Coherences} we outline a procedure
how one could in principle experimentally measure $G_{ij}$, following an idea in \cite{Gullans2019Entanglement}.

The three conditions sufficient for the link with free probability concern the statistical properties of coherences w.r.t.\ to the noise expectation value $\mathbb E_t$ in the limit $1\ll\ell\ll L$. Here we adopt the view that $G_{ij}$ is a random variable with time dependent probability distribution, hence the subscript $t$. 
\begin{enumerate}
\item Local $U(1)$-invariance: The expectation value is invariant under $G_{ij}\to e^{-i\theta_i}G_{ij}e^{i\theta_j}$ which is a multiplication with local phases. In other words, unless $\{i_1,\dots,i_n\}$ is a permutation of $\{j_n,\dots,i_n\}$ we have
\begin{align}\label{eq:u(1)-invariance}
\mathbb E_t[G_{i_1 j_1}\dots G_{i_n j_n}]=0
\end{align}
\item 
Expectation values of "loops" with distinct indices, $i_k\neq i_l, \, \forall k\neq l\in\{1,\cdots,n\}$, scale as
\begin{equation}\label{eq:scaling}
\mathbb E_t [G_{i_1 i_2}G_{i_2 i_3}\dots G_{i_n i_1}] \sim L^{-n+1}
\end{equation}
\item 
Expectation values of products of "loops" factorise
\begin{equation}\label{eq:factorisation_of_loops}
\begin{aligned}&\mathbb E_t [G_{i_1 i_2}\dots G_{i_{m} i_1}G_{j_{1} j_{2}}\dots G_{j_n j_{1}}] \\
&=\mathbb E_t [G_{i_1 i_2}\dots G_{i_{m} i_1}]\mathbb E_t [G_{j_{1} j_{2}}\dots G_{j_{n} j_{1}}]
\end{aligned}
\end{equation}
at leading order. This holds in particular if $i_1=j_1$.
\end{enumerate} 

We speak about "loops" because we can associate diagrams to the expectation values of the various products of $G_{ij}$ by connecting nodes $i$ and $j$ by a directed edge. The diagram associated to (\ref{eq:scaling}) is indeed a loop,
\begin{center}
\includegraphics{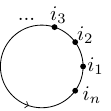}.
\end{center}
In fact, the local $U(1)$ invariance implies that a diagram built in this way is non-zero only if at each node the number of incoming edges is equal to the number of outgoing edges.

For later use we note that the factorisation \eqref{eq:factorisation_of_loops} also implies that connected expectation values scale as
\begin{equation}\label{eq:scaling_connected}
\mathbb E_t [G_{i_1 i_2}G_{i_2 i_3}\dots G_{i_n i_1}]^c=\mathcal O(L^{-n+1})
\end{equation}
even if indices become equal. If for example $i_1=i_k$ and no other indices coincide, then $\mathbb E_t [G_{i_1 i_2}\dots G_{i_{n} i_1}]$ factorises into $\mathbb E_t [G_{i_1 i_2}\dots G_{i_{k-1} i_1}]\mathbb E_t [G_{i_1 i_{k+1}}\dots G_{i_{n} i_1}]$ at leading order. This product of loop expectation values scales as $L^{-(k-1)+1}L^{-(n-k+1)-1}=L^{-n+2}$. The connected expectation value is constructed precisely by subtracting this product of expectation values from $\mathbb E_t [G_{i_1 i_2}\dots G_{i_{n} i_1}]$. What remains, must then scale at least one order of magnitude lower in $L$. This explains \eqref{eq:scaling_connected}. The same argument also shows why the connected expectation value of an arbitrary product of loops can never become more dominant in $L$ than the connected expectation value of single loops of the same length.

As an example consider loops of two points for which \eqref{eq:scaling} implies $\mathbb E[G_{ij}G_{ji}]\sim 1/L$ if $i\neq j$ while \eqref{eq:factorisation_of_loops} implies $\mathbb E[G_{ii}^2]\sim \mathbb E[G_{ii}]^2$ where $\mathbb E[G_{ii}]=\mathcal O(1)$. Therefore $\mathbb E[G_{ij}G_{ji}]^c:=\mathbb E[G_{ij}G_{ji}]-\delta_{ij} \mathbb E[G_{ii}]^2$ scales as most with $1/L$ for any choice of $i$ and $j$. This illustrates nicely that a loops expectation values jump by an order of $L$ if indices become equal, while its connected part does not.

The following subsection gives a definition of the noise average $\mathbb E_t$ and shows how to derive the three conditions from this definition. The consecutive subsection \ref{subsec:link_with_free_probability} shows how the three conditions imply that fluctuations of coherences are in relation with free cumulants from the theory of free probability.

\subsection{Fluctuations in mesoscopic systems}\label{subsec:fluctuations}
We consider a one-dimensional system of spinless interacting fermions without noise on $L$ discrete lattice sites described by a density matrix $\rho$. In order to talk about transport the system is required to have a locally conserved charge that can be transported. In the continuous description of the system, this is the density of particles $n(x,t)$, it satisfies a local conservation law, the continuity equation $\partial_t n+\partial_x j=0$. For the following discussion it is not important if the system is open or closed. 

The separation of time scales between fast and slow degrees of freedom ensures that ballistic cells don't exchange particles with each other during times smaller than the typical time scale $t_\ell$ of ballistic cells. In other words, for $t<t_\ell$, each ballistic cell evolves as a "closed" system and its dynamics is described by a unitary and particle preserving transformations, $\rho\to U^{(i)}\rho\,{U^{(i)}}^\dagger$. Here $U^{(i)}$ acts only on the Hilbert space $\mathcal H^{(i)}\equiv\mathbb C^{(2^\ell)}$ of the cell around $i$. Particle preservation means that
\begin{equation}
[U^{(i)}, N^{(i)}]=0,
\end{equation}
with $N^{(i)}=\sum_{i'\sim i}\hat n_{i'}$ is the total number operator in the cell around site $i$. The sum carries over all sites $i'$ within distance $\ell/2$ from $i$. As a consequence, $U^{(i)}$ decomposes into subspaces $\Lambda_{p}$ ($p$-th exterior product of $\mathbb C^\ell$) of dimension $\ell \choose p$ where the number of fermions is fixed to $p$. That is, $U^{(i)}$ consists of blocks $U^{(i)}_p$ on the diagonal.

To construct the average over ballistic cells, we make the ergodic hypothesis that within a time interval $t_\ell$ the ballistic cell has undergone all possible unitary and particle preserving transformation. Then time averages over intervals $t_\ell$ can be replaced by a uniform Haar average $[\dots]_{ U}$ over such unitaries $U$ that only mix degrees of freedom within a cell, but not between cells.

The idea that fluctuations of a chaotic quantum-many-body system can be characterised through a definition of ergodicity by unitary invariance has already been put forward \cite{Bauer2020Universal}. The consequences of restricting such a global unitary invariance to local sectors of fixed energy has been explored in \cite{Jin2020Equilibration} and \cite{Foini2019Eigenstate} (the latter in the context of ETH, though both talk about very similar things). Here we use similar ideas, but instead of local in energy, we restrict the unitary invariance to be local in space, i.e.\ to unitary invariance within ballistic cells.  

If the separation of two sites $i$ and $j$ is larger than $\ell$, then the average over the corresponding ballistic cells is independent, i.e.\ $ U= U^{(i)}  U^{(j)}$ where $ U^{(i)}$ and $ U^{(j)}$ act only on the cells around $i$ and $j$, respectively. Otherwise, if $i$ and $j$ are in the same cell (but not necessarily equal) we assume that they interact via the same Haar distributed unitary $ U^{(i)}= U^{(j)}$. With this convention, we define the expectation value of coherences to be
\begin{equation}\label{eq:average_G}
\mathbb E_t [G_{ij}]:=\frac 1{t_\ell} \int_t^{t+t_\ell} G_{ij}(t') dt'=\Tr( \rho_t [c_i^\dagger c_j]_{ U}),
\end{equation}
where 
\begin{equation}
[c_i^\dagger c_j]_U:=\int d\mu(U) { U}^\dagger c_i^\dagger c_j  U
\end{equation}
denotes the Haar average. Note that by the cyclic property of the trace, we could have also chosen to average $\rho_t$ instead of $c_i^\dagger c_j$ which makes clear that the average carries over all possible evolutions of the cell.

The time average over $t_\ell$ promotes $G_{ij}(t)$ to a random variable that is only sensitive to the long (with respect to $t_\ell$) time behaviour of the system. Therefore we also expect the right hand side not to depend on the time at which $\rho_{t'}$ is evaluated, as long as $t'\in[t,t+t_\ell]$. Here we chose to evaluate at time $t$.

For quadratic fluctuations this generalises to
\begin{equation}\label{eq:fluctuation_G}
\begin{aligned}
\mathbb E_t [G_{ij}G_{kl}]:=&\frac 1{t_\ell} \int_t^{t+t_\ell} G_{ij}(t') G_{kl}(t') dt' \\
=&\Tr( \rho_t\otimes \rho_t \cdot [c_i^\dagger c_j \otimes c_k^\dagger c_l]_{ U\otimes  U})
\end{aligned}
\end{equation}
and similarly for higher order fluctuations.

\begin{figure}\label{fig:coarse-graining}
\centering
\includegraphics{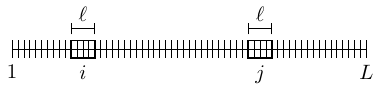}
\caption{The cells around sites $i$ and $j$ in the mesoscopic system above correspond to single sites in the coarse-grained description. The toy model Q-SSEP seems to be a good candidate for such a coarse-grained description. Noise emerges by averaging the mesoscopic system over all unitary transformations that only act on the individual cells and conserves the number of particles inside each cell.}
\end{figure}

\paragraph{U(1)-invariance.}
The averages appearing in \eqref{eq:average_G} and \eqref{eq:fluctuation_G} are invariant under unitary transformations of ballistic cells by construction. For example, we could be interested in an average such as $[c_i^\dagger \otimes c_i]_{ U\otimes  U}$ and unitary invariance means
\begin{equation}
[c_i^\dagger \otimes c_i]_{ U\otimes  U}=V^{(i)}\otimes V^{(i)} [c_i^\dagger \otimes c_i]_{U\otimes  U}{V^{(i)}}^\dagger \otimes {V^{(i)}}^\dagger,
\end{equation}
where $V^{(i)}$ can be any particle preserving unitary on the cell around $i$. In particular, we can choose this $V^{(i)}$ to belong to the subgroup of local $U(1)$ transformations on each site, $V=\exp(\sum_i \theta_i \hat n_i)$, which are generated by the particle number operator $\hat n_i=c_i^\dagger c_i$.  The creation and annihilation operators $c_i^\dagger$ and $c_i$ carry the charge $+1$ and $-1$ with respect to this $U(1)$ transformation. That is $[\hat n_i,c_i^\dagger]=c_i^\dagger$ and  $[\hat n_i,c_i]=-c_i$. Then it is easy to see that any combination of operators inside an average $[\cdots]_{U^{\otimes \cdots}}$ must be composed of an equal number of $c_i^\dagger$'s and $c_i$'s in order for the total charge to add up to zero -- and as a consequence to be invariant under the subgroup of local $U(1)$ transformations. But this is exactly the same as what we claim in \eqref{eq:u(1)-invariance}. For example, \eqref{eq:fluctuation_G} is non-zero only if $i=l$ and $j=k$.

Evaluating averages such as \eqref{eq:fluctuation_G} can be done using Schur's Lemma, but this does not add a new physical insight to the picture. If we restrict ourselves to the one-particle sector, however, one can learn that the local unitary average is equal to a homogeneous sum of $G_{ij}$'s over all sites in the cells. This is shown explicitly in the remainder of this subsection. The discussion can be skipped without impact on the comprehension of the following. 

Unitaries in the one-particle sector are of the form 
\begin{equation}
 U^{(i)} = \exp({\mathbf c^{(i)}}^\dagger M^{(i)} \mathbf c^{(i)}),
\end{equation}
where $\mathbf c^{(i)} = (c_{i-\ell/2},...,c_{i+\ell/2})$ comprises all fermionic annihilation operators inside the cell around $i$, and $M^{(i)}$ is an anti-hermitian $\ell\times\ell$ matrix. Using the identity $e^{-\mathbf c^\dagger M \mathbf c}\,c_i\, e^{\mathbf c^\dagger M \mathbf c} =\sum_j (e^M)_{ij}\,c_j$, where $\mathbf c=(c_1,...,c_L)$ and the definition $ U=e^{\mathbf c^\dagger M \mathbf c}$, where the $L\times L$ matrix $M$ is composed of the blocks $M^{(i)}$ and $M^{(j)}$ on the diagonal at positions $i$ and $j$, and zeros otherwise, this leads to
\begin{equation}\label{eq:simplification_unitary}
\Tr(\rho_t  U^\dagger c_i^\dagger c_j U)=(uG(t)u^\dagger)_{ij}.
\end{equation}
 Here $u:=e^{M^*}=\mathrm{diag}(1,u^{(i)},1,u^{(j)},1)$ is a block diagonal matrix with two unitary blocks $u^{(i)}=e^{M^{(i)*}}$ and $u^{(j)}=e^{M^{(j)*}}$at positions $i$ and $j$ over which the Haar average can be taken (comment \footnote{It is a bit unsatisfying that $U$ depends on the indices of the specific average we consider (here it would be $\mathbb E_t[G_{ij}]$). Intuitively, a general $u$ should be a unitary band matrix with bandwidth of the order of $\ell$. However, since band matrices do not form a group (the bandwidth grows under multiplication) one cannot take Haar averages on them and the construction of the noise would not be well defined.}).
For example, evaluating (\ref{eq:average_G}) using (\ref{eq:simplification_unitary}),
\begin{align}\label{eq:average_G_eval}
\mathbb E_t[G_{ij}]=\sum_{\substack{i'\sim i \\ j'\sim j}}[u^{(i)}_{ii'}G_{i'j'} {u^{(j)}_{jj'}}^*] =
\begin{cases}
\frac 1 \ell \sum_{i'}G_{i'i'} \text{ if } j=i \\
0 \text{ otherwise},
\end{cases}
\end{align}
where the sums carry over all $i'$ and $j'$ in the cells around $i$ and $j$. Note that although expressed through $G_{i'i'}$, the right-hand-side becomes non-random as a result of the law of large numbers.
Evaluating (\ref{eq:fluctuation_G}), using appropriate Haar averages one finds for any choice of $i$ and $j$, up to corrections in $1/\ell$, one finds
\begin{equation}\label{eq:one-particle-fluctuations}
\mathbb E_t[G_{ij} G_{ji}]=\frac 1 {\ell^2} \sum_{\substack{i'\sim i\\j'\sim j}}G_{i'j'}G_{j'i'}+\delta_{ij}(\frac 1 \ell \sum_{i'\sim i}G_{i'i'})^2.
\end{equation}

\paragraph{Scaling with system-size.}
The scaling of loop expectation values \eqref{eq:scaling} with system-size $L$ can be derived from the assumption that the system satisfies the classical macroscopic fluctuation theory (MFT). In particular, this means, that the particle density $n_i$ at site $i$ satisfies a large deviation principle, $\text{Prob}(n_{1},\cdots,n_{L})\sim e^{-L\mathcal F(n_1,\cdots,n_L)}$, where $\mathcal F$ is the so-called quasi potential \cite{Bertini2015MFT}. For a system in equilibrium with a heat bath, $\mathcal F$ would be the difference in free energy between the density profile $\{n_1,\cdots,n_L\}$ in question and the equilibrium density profile. One of the great insights of MFT is that the large deviation principle also extends out-of-equilibrium systems if they are diffusive.

Denoting the noise average at time $t$ within classical MFT by $\langle \cdots \rangle_t$ the large deviation principle implies that connected density correlations scale as
\begin{equation}
\langle n_{i_1} \cdots n_{i_n} \rangle_t^c \sim L^{-n+1},
\end{equation}
where all indices are different.

In the quantum description, we replace $n_i\to \hat n_i = c_i^\dagger c_i$ and $\langle \cdots \rangle_t \to \mathbb E [\Tr(\rho_t\cdots)]$ such that
\begin{equation}\label{eq:density_classical_quantum}
\langle n_{i_1} \cdots n_{i_n} \rangle_t = \mathbb E[\Tr(\rho_t c_{i_1}^\dagger c_{i_1}\cdots c_{i_n}^\dagger c_{i_n})], 
\end{equation}
where $\mathbb E$ is the average over ballistic cells introduced above. If we assume the Hamiltonian to be of the form $H=H_0 + V$, where $H_0$ is quadratic, and furthermore we initialise the system in a Gaussian state of the form $\rho_0=\frac 1 {Z_0} e^{\mathbf c^\dagger M \mathbf c}$ with $Z_0=\Tr(^{\mathbf c^\dagger M \mathbf c})$, then we can apply perturbation theory to decompose \eqref{eq:density_classical_quantum} into a sum of Feynman diagrams.

More precisely, we employ the interaction picture where one divides the full time evolution $U_t=W_t U_t^{(0)}$ into a quadratic evolution $U_t^{(0)}=e^{-iH_0t}$, followed by an evolution with the time ordered exponential $$W_t=T\,\exp(-i\int_0^t dt' V_I(t'))$$ where $V_I(t)=U_t^{(0)}V{U_t^{(0)}}^\dagger$. The density matrix at time $t$ is $$\rho_t=U_t\rho_0U_t^\dagger=W_t\rho_t^{(0)}W_t^\dagger$$ where $\rho_t^{(0)}=U_t^{(0)}\rho_0 {U_t^{(0)}}^\dagger$ preserves its Gaussian form. Now \eqref{eq:density_classical_quantum} becomes
\begin{equation}
\Tr(\rho_t^{(0)} W_t^\dagger c_{i_1}^\dagger c_{i_1}\cdots c_{i_n}^\dagger c_{i_n}W_t)
\end{equation}
which can be expanded using Wick's theorem and the resulting contractions can be denoted by Feynman diagrams.

For $n=2$ we would get
\begin{align}\label{eq:feynman_diagrams}
\Tr(\rho_t c_{i}^\dagger c_{i} c_{j}^\dagger c_{j})
=
\raisebox{-0.5\height}{\includegraphics{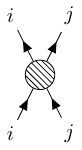}}
-
\raisebox{-0.5\height}{\includegraphics{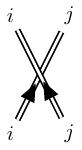}}
+
\raisebox{-0.5\height}{\includegraphics{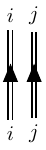}},
\end{align}
where the first diagram is the 1-particle irreducible scattering of two fermions created and annihilated at sites $i$ and $j$, the second diagram is the multiplication of two propagators (Green's functions) between sites $i$ and $j$, i.e.\ $G_{ij}G_{ji}=\Tr(\rho_t c_i^\dagger c_j)\Tr(\rho_t c_j^\dagger c_i)$, and the last diagram is the amplitude to stay on the sites $i$ and $j$, i.e.\ the multiplication of the densities  $G_{ii}G_{jj}$.

When taking the expectation value $\mathbb E$, these three diagrams should be equal to the corresponding value within MFT,
\begin{equation}
\langle n_i n_j \rangle_t = \langle n_i n_j \rangle^c_t +\langle n_i \rangle_t \langle n_j \rangle_t.
\end{equation}
Since the second term $\langle n_i \rangle_t \langle n_j \rangle_t=\mathbb E_t[G_{ii}] \mathbb E_t[G_{jj}]$ (to leading order) is equal to the third diagram, the first term $\langle n_i n_j \rangle^c_t\sim L^{-1}$ must equal the first plus the second diagram in \eqref{eq:feynman_diagrams}. If we assume that the expectation value $\mathbb E$ of these two diagrams has the same scaling with $L$, we can conclude that the expectation value of the second diagram satisfies the scaling we claimed in the beginning of the section, that is $\mathbb E[G_{ij}G_{ji}]\sim L^{-1}$. This argument can be repeated for any $n$ in a similar fashion if one assumes that the expectation value of 1-particle irreducible diagrams with more legs has the same scaling as the expectation values of the multiplication of several crossing propagators.
We admit that this assumption is a weak point in our argument and one should start here if one is interested to understand for which class of Hamiltonians this general picture applies. In particular, our argument would not apply if perturbation theory breaks down, signalling a possible phase transition, or in presence of elementary excitation of a non-perturbative nature.

\paragraph{Factorisation of products of loops.}
The factorisation in \eqref{eq:factorisation_of_loops} for the case $i_1\neq j_1$ is trivial, because the average $\mathbb E_t$ treats each ballistic cell independently by construction. It remains to treat the case $i_1=j_1$. We illustrate this case by an example of a $2$-loop.

Assuming the scaling from the last section, a look at \eqref{eq:one-particle-fluctuations} reveals that for $i=j$ the leading contribution comes from the second terms which is $\mathcal O(1)$ while the first terms is $\mathcal O(1/L)$ (comment \footnote{There is a small subtlety: The first term consists of $\ell$ terms of the form $G_{i'i'}^2$ when $i'=j'$. These also scale as $\mathcal O(1)$, but are suppressed by the $1/\ell^2$-factor. Their total contribution is $1/\ell$ which goes to zero}). That is, to leading order, and restricted to the one-particle sector, we have,
\begin{equation}\label{eq:one-particle-factorization}
\mathbb E_t[G_{ii}^2]=(\frac 1 \ell \sum_{i'\sim i}G_{i'i'})^2=\mathbb E_t[G_{ii}]^2.
\end{equation}
In the last equality we used \eqref{eq:average_G_eval}. This calculation shows that for the given example the expectation value of products of loops (here $G_{ii}^2$) is indeed proportional to the product of the expectation values of loops (here single loops $G_{ii}$) at leading order. Using Schur's Lemma, one can show that this statement remains true at all particle sectors and for higher order fluctuations, thereby confirming \eqref{eq:factorisation_of_loops}.

\subsection{Link with free probability}\label{subsec:link_with_free_probability}
The link with free probability occurs when expressing loop expectation values $\mathbb E[G_{i_1i_2}...G_{i_ni_1}]$ in terms of their connected parts. We will find that this expansion involves a sum over non-crossing partitions only, see \eqref{eq:moment-cumulant_Q-SSEP}. Such non-crossing partitions are at the heart of a mathematical theory of non-commuting random variables that is called \textit{free probability theory} and introduced in the next section. By means of non-crossing partitions one can define so-called \textit{free cumulants} which will arise naturally in our setting.
 
As an intermediate step and as usual in random matrix theory, we consider the measure of $G$, the matrix of coherences $G_{ij}$, defined by the expectation value of its traces,
\begin{equation}\label{eq:moments_G}
	\phi(G^n):=\frac 1 L \mathbb E[\Tr(G^n)]=\frac 1 L\sum_{i_1,\cdots,i_n} \mathbb E[G_{i_1i_2}\cdots G_{i_ni_1}].
\end{equation}
Here we dropped the subscript $t$ of the measure $\mathbb E_t$ for simplicity. Whenever two indices $i_k$ in the above sum are equal, the expectation value factorises according to (\ref{eq:factorisation_of_loops}). We therefore split this sum into sums where all indices are distinct -- except for a given set of indices that are imposed to be equal,
\begin{equation}
\begin{aligned}
\sum_{i_1,\cdots,i_n}=\sum_{\substack{i_1,\cdots,i_n \\ \text{distinct}}}+\sum_{\substack{i_1=i_2,i_3\cdots,i_n \\ \text{distinct}}}+\cdots+\sum_{i_1=i_2=\cdots=i_n}
\end{aligned}
\end{equation}
Such a splitting can be understood as a sum over partitions $\pi\in P(n)$ of the set $\{1,\cdots,n\}$ into blocks $b\in\pi$ that group together all the indices $i_k$ that are imposed to be equal. For example, the partitions corresponding to the three terms above are $\pi=\{\{1\},\cdots,\{n\}\}$, $\pi=\{\{1,2\},\{3\},\cdots,\{n\}\}$ and $\pi=\{\{1,2,\cdots,n\}\}$. The sum becomes
\begin{equation}
\sum_{i_1,...,i_n}=\sum_{\pi\in P(n)} \sum_{\substack{i_1,...,i_n \text{distinct,} \\ \text{except }i_k=i_l \text{ whenever } k,l \\ \text{are in the same block of }\pi}}.
\end{equation}
For $n=4$ two possible partitions are for example
\begin{align}\label{eq:pi1}
\pi_1=\{\{1,3\},\{2\},\{4\}\}=
\raisebox{-0.5\height}{\includegraphics{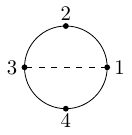}}
\end{align}
and 
\begin{align}\label{eq:pi2}
\pi_2=\{\{1,3\},\{2,4\}\}=
\raisebox{-0.5\height}{\includegraphics{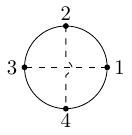}}
\end{align}
Representing partitions by diagrams where nodes in the same block are connected by dashed lines shows very intuitively that $\pi_2$ is a crossing partition while $\pi_1$ is non-crossing. It turns out that in the sum (\ref{eq:moments_G}) terms corresponding to crossing partitions are of the order of $1/L$ and lower and therefore vanish for $L\to\infty$, while all non-crossing partitions are of order one. Here we illustrate this fact for $n=4$ and the two examples above. The term corresponding to $\pi_1$ factorises and becomes
\begin{align}
\frac 1 L \sum_{\substack{i_1=i_3,i_2,i_4 \\ \text{distinct}}} \mathbb E[G_{i_1i_2}G_{i_2i_1}]\mathbb E[G_{i_3i_4} G_{i_4i_3}] = \mathcal O(1).
\end{align}
Each of the $2$-loop expectation values scale as $1/L$ and each term in the sum is of order $1/L^3$.  Since the sum carries over three indices running from $1$ to $L$ this cancels and the resulting scaling is of order one. 

In contrast to this the term corresponding to $\pi_2$ can factorise in two different ways and becomes
\begin{align}
\frac 1 L \sum_{\substack{i_1,i_2 \\ \text{distinct}}} 2\, \mathbb E[G_{i_1i_2}G_{i_2i_1}]^2= \mathcal O(1/L).
\end{align}
The difference to $\pi_1$ is that now there are only two indices to sum over and hence the scaling is of order $1/L$.

For the surviving non-crossing partitions one can ask how the partition, that determines which indices $i_k$ are equal, is related to the product of loop expectation values of $G_{ij}$'s that appears after the factorisation. If we modify (\ref{eq:pi1}) by connecting as many edges by solid lines as possible without crossing a dashed line, i.e.
\begin{equation}
	\raisebox{-0.5\height}{\includegraphics{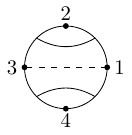}},
\end{equation}
we see that for the partition $\pi_1$ this product of loop expectation values, i.e.\ $\mathbb E[G_{i_1i_2}G_{i_2i_1}]\mathbb E[G_{i_3i_4} G_{i_4i_3}]$, is determined by the solid lines. In fact, for every non-crossing partition $\pi$ of nodes, the solid lines define a unique non-crossing partition $\pi^*$ of edges, the \textit{dual partition} (also called \textit{Kreweras complement}). If we label an edge by the adjacent node with the lower number we have for example $\pi_1^*=\{\{1,2\},\{3,4\}\}$. Each block therein corresponds to a loop of $G_{ij}$'s. In this terminology the expansion of $\phi(G^n)$ in (\ref{eq:moments_G}) becomes
\begin{equation}
\frac 1 L \sum_{\pi\in NC(n)} \sum_{\substack{i_1,...,i_n \\ \text{distinct,} \\ \text{except }i_k=i_l \\ \text{ if } k,l\in b \text{ for }b\in\pi}} \prod_{b\in\pi^*} \mathbb E[G_{i_{b(1)}i_{b(2)}}...G_{i_{b(|b|)}i_{b(1)}}],
\end{equation}
where $NC(n)$ denotes all non-crossing partitions of the set $\{1,\cdots,n\}$.

Instead of using this slightly akward sum over indices $i_k$ we introduce a modified Kronecker-delta 
\begin{equation}\label{eq:kronecker-delta_pi}
\delta_\pi\equiv\delta_\pi(i_1,\cdots,i_n)=\prod_{b\in\pi} \delta_{i_{b(1)},\cdots,i_{b(|b|)}}
\end{equation}
that sets all indices to be equal that belong to the same block in $\pi$. Next, we extend the sum to include the cases in which indices $i_k$ can become equal by replacing $\mathbb E[\cdots]$ with $\mathbb E[\cdots]^c$. This introduces only a sub-leading error of order $1/L$, because \eqref{eq:scaling_connected} ensures that connected expectation values do not get more dominant when some indices become equal.

Then the expansion of $\phi(G^n)$ in (\ref{eq:moments_G}) reads
\begin{equation}\label{eq:moments_G_expansion}
\frac 1 L \sum_{\pi\in NC(n)} \sum_{i_1,...,i_n} \delta_\pi \prod_{b\in\pi^*} \mathbb E[G_{i_{b(1)}i_{b(2)}}...G_{i_{b(|b|)}i_{b(1)}}]^c.
\end{equation}
In free probability $\phi$ corresponds to the "expectation" value of the random matrix $G$ and we have shown that the moments of $G$ have a natural expansion in terms of a sum of non-crossing partitions. The terms in this sum are usually called free cumulants. However, comparing to the definition of free cumulants \eqref{eq:def_free_cumulants} in the next section, one sees that the expansion above has a slightly different form due to the additional $\delta_\pi$. This issue is further discussed in section \ref{subsec:steady_state_solution}

To conclude the argument, note that we could have done the whole derivation multiplying $\mathbb E[G_{i_1i_2}\cdots G_{i_ni_1}]$ in \eqref{eq:moments_G} with a smooth test function $h_{i_1,\cdots,i_n}$ that would again appear in (\ref{eq:moments_G_expansion}). By comparing the two equations we would find that
\begin{equation}\label{eq:moment-cumulant_Q-SSEP}
\begin{aligned}
&\mathbb E[G_{i_1i_2}...G_{i_ni_1}]
\\
&=\sum_{\pi\in NC(n)}\mathbb \delta_{\pi^*}(i_1,...,i_n)\prod_{b\in \pi}\mathbb E[G_{i_{b(1)}i_{b(2)}}...G_{i_{b(|b|)}i_{b(1)}}]^c,
\end{aligned}
\end{equation}
an equation we will reuse in section \ref{sec:Q-SSEP}. Here we interchanged the role of $\pi$ and $\pi^*$ which is possible because, for non-crossing partitions, they are in one-to-one correspondence. A full proof of this formula for any $n$ is given in appendix \ref{app:non-crossing_partitions}.

\subsection{Analogy with ETH}
Simultaneously to our work it has been observed in \cite{Pappalardi2022ETH} that there is a very similar link to free probability in the context of the eigenstate thermalisation hypothesis (ETH). Indeed, on the mathematical level, fluctuations of spatial coherences $G_{ij}$ in one dimensional mesoscopic systems seem to behave in complete analogy to matrix elements $A_{ij}=\langle E_i|A|E_j\rangle$ of a local observable $A$ expressed in the energy eigenbasis $|E_i\rangle$ of a hamiltonian $H$ in a closed system that obeys ETH. In the context of ETH, $A_{ij}$ is a random variable with respect to an fictitious ETH-random-matrix-ensemble that captures its typical behaviour. 

Comparing to \cite{Foini2019Eigenstate}, which discusses higher order fluctuation of $A_{ij}$ within ETH, one realises that the emergence of the three properties \eqref{eq:u(1)-invariance}-\eqref{eq:factorisation_of_loops} have analogous reasons. In ETH they are the result of an average over small energy windows. In our context they are the result of an average over small space windows, which we called ballistic cells. To complete the analogy, loop expectation values of $A_{ij}$ in ETH scale with the density of states $e^{S(E_+)}$ at energy $E_+=\frac 1 2 (E_i+E_j)$. In our context, this corresponds to the "density of states" at sites $i$ and $j$, which is just a constant and equal to the number of sites $L$ (because the physical length of our system is set to one for simplicity).

\section{Introduction to free probability}\label{sec:free_probability}

In classical probability, two variables are independent if (and only if) their
moments factorise at all orders, $\mathbb{E}[X^{n}Y^{m}]=\mathbb{E}[X^{n}]\mathbb{E}[Y^{m}]$
for all $n,m\in\mathbb{N}$. One can therefore determine joint moments of any product of independent variables knowing only the moments of the individual independent variables. If instead $X$ and $Y$ are random matrices with independent entries, then it is less clear how we would achieve the factorisation of, say, $\mathbb E[XYXY]$ into the moments of the independent variables $\mathbb E[X^2]$ and $\mathbb E[Y^2]$, on the level of matrices, since they don't commute in general. 

Free probability theory solves this issue by proposing an extension of the notion of independence for non-commutative random variables, called \textit{freeness}. \textit{Free} variables are not only required to be independent in the probabilistic sense, but also to be algebraically independent, in the sense that there are no algebraic relations between the variables. This is similarly to generators in a free group, hence the name.

Given two non-commuting random variables $a$ and $b$ in some algebra $\mathcal{M}$ (e.g.\ algebra of large random matrices) and a linear functional $\varphi:\mathcal{M}\to\mathbb{C}$ (that plays the role of the expectation value), then $a$ and $b$ are called \textit{free} if for all polynomials $P_1,\cdots,P_l$ and $Q_1,\cdots,Q_l$ with $\varphi(P_i(a))=0$ and $\varphi(Q_i(b))=0$ we have
\begin{equation}
\varphi(P_1(a)Q_1(b)\cdots P_l(a) Q_l(b))=0.
\end{equation}
The reason to evoke all possible polynomials in the definition is that any element in the subalgebras $\mathcal A$ and $\mathcal B$ generated by $a$ and $b$, respectively, can be written as $P_i(a)$ and $Q_i(b)$. Hence freeness can also be understood as a statement about the subalgebras $\mathcal A$ and $\mathcal B$.

This definition implies for example that if $a$ and $b$ are free, then $\varphi(a^{m}b^{n})=\varphi(a^{m})\varphi(b^{n})$.
Additional structure occurs if one interchanges the order
such that free variables are no longer grouped together. But it remains always true that joint moments can be determined through the moments of the individual free variables therein. For example
$\varphi(abab)=\varphi(a^2)\varphi(b)^2+\varphi(a)^2\varphi(b^2)-\varphi(a)^2\varphi(b)^2$.

The definition of freeness was proposed by Dan Voiculescu in 1985, who founded the field of free probability theory while working on problems in operator algebras. More details can be found in his book \cite{Voiculescu1997Free} and a good introduction to the subject provide the lecture notes by Roland Speicher \cite{Speicher2019Lecture} as well as the book by Mingo and Speicher \cite{Mingo2017Free}.

In the 1990's, Speicher proposed a complementary combinatorial approach to free probability by introducing what he called \textit{free cumulants}. This is also the route we took in the last section to make a connection between coherent fluctuations and free probability theory. Before introducing free cumulants, let us review how cumulants in classical probability theory are defined and how they are related to partitions. This will allow us then to better appreciate why free cumulants are defined via non-crossing partitions. 

\subsection{Classical cumulants and partitions\label{subsec:Standard-cumulants}}

Let $\{X_{1},\cdots,X_{N}\}$ be a family of classical random variables
with moment-generating-function
\begin{align*}
Z[a,u]  &:=\mathbb{E}[e^{u\sum_{i}a_{i}X_{i}}]\\
&=\sum_{n\ge0}\frac{u^{n}}{n!}\sum_{i_{1}\cdots i_{n}}a_{i_{1}}\cdots a_{i_{n}}\mathbb{E}[X_{i_{1}}\cdots X_{i_{n}}],
\end{align*}
where $u$ is an (optional) parameter that counts the order of the joint
moment (or correlation function) $\mathbb{E}[X_{i_{1}}\cdots X_{i_{n}}]$. The joint cumulant (or connected correlation function) $\mathbb{E}[X_{i_{1}}\cdots X_{i_{n}}]^{c}$ is defined as the term proportional to $a_{i_{1}}\cdots a_{i_{n}}$ in
the expansion of the cumulant generating function $W[a,u]:=\log Z[a,u]$,
\[
W[a,u]=\sum_{n\ge0}\frac{u^{n}}{n!}\sum_{i_{1}\cdots i_{n}}a_{i_{1}}\cdots a_{i_{n}}\mathbb{E}[X_{i_{1}},\cdots, X_{i_{n}}]^{c}.
\]
In fact, cumulants and moments are related by a combinatorial formula.
Expanding $Z[a,u]$ in terms of the cumulants and grouping together
terms with the same power of $u$ one can derive that a moment $\mathbb{E}[X_{i_{1}}\cdots X_{i_{n}}]$
can be expressed as a sum of products of cumulants arranged according
to partitions $\pi$ of the set $\{i_{1},\cdots,i_{n}\}$,
\begin{equation}\label{eq:moments_as_cumulants}
\mathbb{E}[X_{i_{1}}\cdots X_{i_{n}}]=\sum_{\pi\in P(n)}\prod_{b\in\pi}\mathbb{E}[X_{i_{b(1)}}X_{i_{b(2)}}\cdots]^{c},
\end{equation}
where $b=\{b(1),b(2),\cdots\}$ denotes the elements of a block of the
partition $\pi$. The number of partitions of a set of $n$ elements is called the Bell number $B_n$, with recursion relation $B_{n+1}=\sum_{k=0}^n \big({}^n_k\big)B_k$ and $B_1 = 1$, $B_2 = 2$, $B_3 = 5$, $B_4 = 15$ and $B_5 = 52$, etc.

Let us give an example for $n=4$ and the set $\{1,2,3,4\}$. For example, the partition $\pi=\{\{1,2\},\{3,4\}\}$ is represented by the diagram
\begin{equation}
	\raisebox{-0.5\height}{\includegraphics{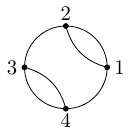}}
\end{equation}
This is in full analogy to (\ref{eq:pi1}) and (\ref{eq:pi2}), except that here we used solid lines for better visibility, since we don't talk about dual partitions in this section. The expansion of the moment $\mathbb{E}[X_{1}X_{2}X_{3}X_{4}]$ in terms
of products of cumulants $\prod_{b\in\pi}\mathbb{E}[X_{i_{b(1)}}X_{i_{b(2)}}\cdots]^{c}$, represented through the diagram of the corresponding partition $\pi$, becomes
\begin{align}\label{eq:moment_cumulant_4}
	&\raisebox{-0.5\height}{\includegraphics{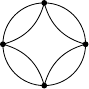}}
+
	\raisebox{-0.5\height}{\includegraphics{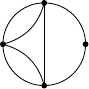}}_{\circlearrowleft4} \hskip -0.2 truecm
+
	\raisebox{-0.5\height}{\includegraphics{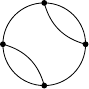}}_{\circlearrowleft2} \hskip -0.2 truecm
+
	\raisebox{-0.5\height}{\includegraphics{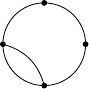}}_{\circlearrowleft4}\nonumber \\
&+
	\raisebox{-0.5\height}{\includegraphics{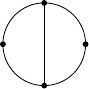}}_{\circlearrowleft2} \hskip -0.2 truecm
+
	\raisebox{-0.5\height}{\includegraphics{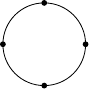}}
+
	\raisebox{-0.5\height}{\includegraphics{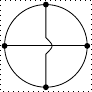}}
\end{align}

where the subscript ${\circlearrowleft\!k}$ suggests that by cyclic
permutation there are in total $k$ such diagram. Note the fact, that
the last diagram (in a dotted box) corresponds to a crossing partition
$\pi=\{\{1,3\},\{2,4\}\}$. In free probability theory these diagrams
do not appear as we will see below.

\paragraph{Classical cumulants of a single variable.}
The moment-cumulant relation \eqref{eq:moments_as_cumulants} allows
us to express the cumulants recursively through the moments. In the
case of a single variable $X=X_{1}=\cdots=X_{N}$ we illustrate how this
can be done up to order four. Let us denote by $m_{n}=\mathbb{E}[X^{n}]$
and $c_{n}=\mathbb{E}[X^{n}]^{c}$ the moments and cumulants of this
variable, then
\begin{align}
m_{1} & =c_{1},\label{eq:moment_cumulants_single_variable}\\
m_{2} & =c_{2}+c_{1}^{2},\nonumber \\
m_{3} & =c_{3}+3c_{2}c_{1}+c_{1}^{3},\nonumber \\
m_{4} & =c_{4}+4c_{1}c_{3}+3c_{2}^{2}+6c_{1}^{2}c_{2}+c_{1}^{4}.\nonumber 
\end{align}
Note that the coefficients correspond exactly to the cyclic multiplicities of the diagrams. This can be solved recursively for $c_{k}$, 
\begin{align}
c_{1} & =m_{1},\label{eq:free_cumulants_moments_single_variable}\\
c_{2} & =m_{2}-m_{1}^{2},\nonumber \\
c_{3} & =m_{3}-3m_{1}m_{2}+2m_{1}^{3},\nonumber \\
c_{4} & =m_{4}-4m_{1}m_{3}+12 m_{1}^{2}m_{2}-3m_{2}^{2}-6 m_{1}^{4}.\nonumber 
\end{align}
The generating function of cumulants is the logarithm of that of moments, as explained above.
Below we will see how this formula differs for the free cumulants

\subsection{Free cumulants and non-crossing partitions\label{subsec:free_cumulants}}
In the setting of non-commuting random variables $a_1,\cdots,a_N$ in some algebra $\mathcal M$ with "expectation value" $\varphi\colon\mathcal M\to\mathbb C$, the free cumulants $\kappa_n$ are multilinear forms implicitly defined  through moments by a sum over non-crossing partitions $\pi\in NC(n)$ of the set $\{i_{1},\cdots,i_{n}\}$,
\begin{equation}\label{eq:def_free_cumulants}
\varphi(a_{i_{1}}\cdots a_{i_{n}})=:\sum_{\pi\in NC(n)}\prod_{b\in\pi}\kappa_{|b|}(a_{i_{b(1)}},a_{i_{b(2)}},\cdots),
\end{equation}
where $|b|$ denotes the number of elements in the block $b=\{b(1),b(2),\cdots\}$. For example, if we would expand $\varphi(a_{1}a_{2}a_{3}a_{4})$ in terms of free cumulants we could get all diagrams in \eqref{eq:moment_cumulant_4}, except the last one which is crossing.  Therefore the order of the arguments in $\kappa_n$ becomes important, even if the $a_i$ were to commute, hence the separation by the comma.

The free cumulants satisfy a number of properties that are analogous to properties of classical cumulants:
\begin{itemize}
\item $\kappa_n(a_1,\cdots,a_n)=0$ as soon as there appears a pair $a_i,a_k$ of free variables, see e.g.\ \cite{Speicher2019Lecture}
\item As a result of multiliniarity and the last bullet point, free cumulants of free variables $a$ and $b$ are additive, $\kappa_n(a+b,\cdots,a+b)=\kappa_n(a,\cdots,a)+\kappa_n(b,\cdots,b)$, see \cite{Novak2011AMS}
\item Any variable $a$ whose free cumulants $\kappa_n(a,\cdots,a)$ vanish for $n\ge3$ is distributed according Wigner's semi-circle law of random matrix theory for the Gaussian unitary ensemble (GUE), so that large GUE random matrices are analogous of Gaussian variables but from a free probability point of view. 
\end{itemize}

The number of non-crossing partitions of a set of $n$ elements is the Catalan number $C_n=\frac{1}{n+1}\big({}^{2n}_n\big)$, with $C_1=1$, $C_2=2$, $C_3=5$, $C_4=14$ and $C_5=42$, etc.
The formula \eqref{eq:def_free_cumulants} is triangular and can be inverted to express the free cumulants in terms of the moments. 

\paragraph{Free cumulants of a single variable.}
In the case of a single variable $a=a_{1}=...=a_{N}$, we denote by $\kappa_{n}:=\kappa_n(a,\cdots,a)$ and $m_{n}:=\varphi(a^{n})$ the free cumulant and the moment at order $n$, respectively. Then we have 
\begin{align}
m_{1} & =\kappa_{1},\label{eq:moments_free_cumulants_4}\\
m_{2} & =\kappa_{2}+\kappa_{1}^{2},\nonumber \\
m_{3} & =\kappa_{3}+3\kappa_{2}\kappa_{1}+\kappa_{1}^{3},\nonumber \\
m_{4} & =\kappa_{4}+4\kappa_{1}\kappa_{3}+2\kappa_{2}^{2}+6\kappa_{1}^{2}\kappa_{2}+\kappa_{1}^{4}.\nonumber 
\end{align}
The equations can be solved for $\kappa_{k}$ recursively, 
\begin{align}
\kappa_{1} & =m_{1},\label{eq:free_cumulants_moments_4}\\
\kappa_{2} & =m_{2}-m_{1}^{2},\nonumber \\
\kappa_{3} & =m_{3}-3m_{1}m_{2}+2m_{1}^{3},\nonumber \\
\kappa_{4} & =m_{4}-4m_{1}m_{3}+10m_{1}^{2}m_{2}-2m_{2}^{2}-5m_{1}^{4}.\nonumber 
\end{align}
Note, that the difference between standard and free cumulants only
shows up at order $4$ since here a crossing-partition become possible
for the first time. For a single variable, the relation between the moments and the free cumulants can be found by inverting the resolvent associated to the distribution, see \cite{Speicher2019Lecture,Biane1998Proba}. 

\subsection{Free cumulants in Random Matrix Theory}

A relation between free probability theory and random matrices was first observed by Voiculescu in 1991 \cite{Voiculescu1991Limit}. For example, he realised that matrices in the Gaussian unitary ensemble (GUE) with independent entries become free variables in the limit where the dimension of the matrix goes to infinity. Since then many more connection between other random matrix ensembles and free probability have been found. 

Here we will make one of these results more explicit, which applies to matrices that are rotated by random unitaries. Consider $N\times N$ random matrices of the form $X_A=U_{N}A_{N}U_{N}^{\dagger}$ and $Y_N=U_{N}B_{N}U_{N}^{\dagger}$,
where $U_{N}$'s are choses independently from the Haar distribution
over the unitary group and $A_{N}$ and $B_N$ are deterministic matrices with spectral densities $\mu_{A}$ and $\mu_B$. That is, the moments $m_{k}:=\lim_{N\to\infty}\frac{1}{N}\mathrm{Tr}(A_{N}^{k})=\int\lambda^{k}\mu_{A}(\lambda)d\lambda$ are all finite, and similarly for $B_N$. The statement is that in the limit $N\to\infty$ the random matrices $X_N$ and $Y_N$ become free variables $a$ and $b$ (in some non-commutative probability space) with distribution $\mu_A$ and $\mu_B$ with respect to the measure $\varphi:=\frac 1 L \mathbb E\, \Tr$, where $\mathbb E$ is the Haar measure.

It is furthermore known that the classical cumulants of such
matrices $X_{N}$ can be expressed as the free cumulants $\kappa_n\equiv\kappa_n(a,\cdots,a)$ of the spectral density $\mu_{A}$. That is,
\begin{equation} \label{eq:large-HCIZ}
   \mathbb{E}[e^{N\mathrm{Tr}(P_{N}X_{N})}]\asymp_{N\to\infty} e^{N\sum_{n=1}\frac{1}{n}\kappa_{n}\mathrm{Tr}(P_N^{n})}, 
\end{equation}
where $P_{N}$ is a sequence of matrices with fixed rank (such that
$\mathrm{Tr}(P_{N}^k)$ does not scale with $N$), for instance a rank one projector.

This connection between Haar distributed random matrices and free cumulants is adapted to the closed Q-SSEP in the steady state -- which actually corresponds to an equilibrium situation due to the absence of current in the steady state. Indeed, since in the closed case the Q-SSEP dynamics is unitary and ergodic over the unitary group, its steady state distribution is the one induced by the Haar measure on the orbit of the initial matrix of coherences $G_0$ \cite{Bauer2019Equilibrium}. That is, the matrix of coherence $G$ is distributed as $UG_0U^\dag$ with $U$ a Haar distributed unitary $L\times L$ matrix. As a consequence of \eqref{eq:large-HCIZ} above, the large deviation function for coherence fluctuations in the closed Q-SSEP is the generating function of the free cumulants of the spectral measure of the initial matrix of coherences $G_0$.

\subsection{Alternative derivation of free probability in mesoscopic systems}\label{subsec:link_with_free_probability_alt}
As an alternative to section \ref{subsec:link_with_free_probability}, one can derivation the link between fluctuations in mesoscopic systems and free probability starting from the moment-cumulant formula \eqref{eq:moments_as_cumulants} introduced earlier in this section. This is also the way we chose in appendix \ref{app:non-crossing_partitions} which contains the full proof of this link for any $n$.

We apply  \eqref{eq:moments_as_cumulants} to a loop of $G_{ij}$'s and obtain 
\begin{equation}\label{eq:moment-cumulant_Gs}
\begin{aligned}
\mathbb E [G_{i_1i_2}...G_{i_ni_1}]\\=\sum_{\pi\in P(n)}\mathbb \prod_{b\in \pi}&\mathbb E[G_{i_{b(1)}i_{b(1)+1}}... G_{i_{b(|b|)}i_{b(|b|)+1}}]^c
\end{aligned}
\end{equation}
where the subscript $t$ of $\mathbb E_t$ is dropped for simplicity.

For $n=4$ two possible partitions, each with two blocks, are
\begin{align}\label{eq:pi1_alt}
\pi_1=\{\{1,2\},\{3,4\}\}&=
	\raisebox{-0.5\height}{\includegraphics{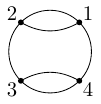}}
\end{align}
and 
\begin{align}
\pi_2&=\{\{1,3\},\{2,4\}\}=
	\raisebox{-0.5\height}{\includegraphics{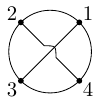}}.
\end{align}
We determine the contribution of the two partitions:
\begin{align}
\pi_1\colon\ \ &\mathbb E[G_{i_1i_2}G_{i_2i_3}]^c\mathbb E[G_{i_3i_4}G_{i_4i_1}]^c& \\ \nonumber
&=\delta_{i_1i_3} \mathbb E[G_{i_1i_2}G_{i_2i_1}]^c\mathbb E[G_{i_3i_4}G_{i_4i_3}]^c = \mathcal O(L^{-3}).
\end{align}
In the second line we used the $U(1)$ invariance of $\mathbb E$ that implies that the expression is non-zero only for $i_1=i_3$. Finally we evaluated the scaling of the expression with $L$. Here we used that $\mathbb E[G_{ij}G_{ji}]^c=\mathcal O(L^{-1})$ by \eqref{eq:scaling_connected} and that the Kronecker delta becomes $\delta_{ij}\to\delta(x-y)/L$ for large $L$, where $x=i/L$ and $y=j/L$. Note that the resulting scaling with $\mathcal O(L^{-3})$ is the same as that of a $4$-loop, which also appears in the sum over partitions as the partition $\{\{1,2,3,4\}\}$ that contains only a single block. It turns out, that all non-crossing partitions have the same scaling with $L$. 

In contrast to this, crossing partitions turn out to be sub-leading. Evaluating $\pi_2$ one has
\begin{align}
\pi_2\colon\ \ &\mathbb E[G_{i_1i_2}G_{i_3i_4}]^c\mathbb E[G_{i_2i_3}G_{i_4i_1}]^c& \\ \nonumber
&=\delta_{i_2i_3}\delta_{i_1i_4} \mathbb E[G_{i_1i_2}G_{i_2i_1}]^c\mathbb E[G_{i_2i_2}G_{i_1i_1}]^c =\mathcal O(L^{-4}).
\end{align}
Here we assumed that no connected expectation value is more dominant than that of loops with the same number of points, i.e. $\mathbb E[G_{i_1i_1}G_{i_2i_2}]^c=\mathcal O( L^{-1})$, because the loop with the same number of points scales as $\mathcal O( L^{-1})$.

Note that the average over independent ballistic cells introduced in section \ref{subsec:fluctuations} actually implies that $\mathbb E[G_{ii}G_{kk}]^c=0$ if $i\neq k$. However, we anticipate that the toy model Q-SSEP predicts this term to scale with $L^{-2}$ if $i\neq k$. A possible conclusion is that even though the average over ballistic cells is a good approximation for leading order terms, it is probably too crude to capture the complete dependence of correlation functions on the system size $L$.

In the case of non-crossing partitions, the delta functions that we introduce each time to "close" the blocks into loops, can be elegantly described via the dual partition (Kreweras complement) $\pi^*$. This is best understood diagrammatically. To $\pi_1$ we associate the diagramm
\begin{equation}
	\raisebox{-0.5\height}{\includegraphics{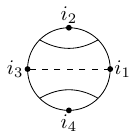}}.
\end{equation}
The solid lines represent blocks in $\pi_1$ (here understood as a partition of the edges) while the dashed lines define the blocks of a partition of the nodes, the dual partition $\pi_1^*=\{\{1,3\},\{2\},\{4\}\}$. If $\pi$ is non-crossing then there is a unique way to connect a maximal number of nodes by dashed lines without crossing a solid line. To a dual partition $\pi^*$ we associate a Kronecker delta $\delta_{\pi^*}(i_1,...,i_n)$, that equates all indices belonging to the same block of $\pi^*$, exactly as in \eqref{eq:kronecker-delta_pi} but with $\pi$ and $\pi^*$ interchanged. For example $\delta_{\pi_1^*}(i_1,...,i_4)=\delta_{i_1i_3}$.

This notation allows us to rewrite the moment-cumulant-formula (\ref{eq:moment-cumulant_Gs}) as a sum over non-crossing partitions only. Denoting the set of non-crossing partitions of $\{1,\dots,n\}$ by $NC(n)$, this is
\begin{equation}\label{eq:moment-cumulant_Q-SSEP_alt}
\begin{aligned}
&\mathbb E [G_{i_1i_2}...G_{i_ni_1}]
\\
&=\sum_{\pi\in NC(n)}\mathbb \delta_{\pi^*}(i_1,...,i_n)\prod_{b\in \pi}\mathbb E[G_{i_{b(1)}i_{b(2)}}...G_{i_{b(|b|)}i_{b(1)}}]^c.
\end{aligned}
\end{equation}
This is exactly the same formula as \eqref{eq:moment-cumulant_Q-SSEP}.

\section{The open quantum SSEP}\label{sec:Q-SSEP}
\subsection{Summary of results}\label{subsec:summary_of_results}
The quantum symmetric simple exclusion process (Q-SSEP) describes a one-dimensional chain with $L$ sites on which
spinless fermions can hop to neighbouring sites with random amplitudes. In the closed Q-SSEP this chain is closed periodically, while in the open Q-SSEP the chain is coupled to two reservoirs that can inject and extract fermions on the boundary with different rates. It is therefore possible to have a steady current of fermions through the chain keeping the system out-of-equilibrium. The model allows to study the time evolution of spatial coherences $G_{ij}(t)=\mathrm{Tr}(\rho_{t}c_{i}^{\dagger}c_{j})$ which fluctuate due to the random hopping. The measure $\mathbb E_t$ of this randomness converges to a unique steady measure $\mathbb E_\infty$ after long times which has been characterised in \cite{Bernard2019Open}. 

The main results on the open Q-SSEP in this paper are equations for the time evolution of the $n$-loop expectation value $\mathbb{E}_{t}[G_{i_{1}i_{2}}G_{i_{2}i_{3}}\cdots G_{i_{n}i_{1}}]$ and its connected part, in a non-trivial scaling limit where $L$ becomes large. In deriving them, we make use of non-crossing partitions which appear in the same way as in section \ref{subsec:link_with_free_probability} because Q-SSEP satisfies the three conditions \eqref{eq:u(1)-invariance}-\eqref{eq:factorisation_of_loops} (see section \ref{subsec:intro_Q-SSEP}).

It has been shown in \cite{Bernard2022DynamicsClosed},
that the closed Q-SSEP has a meaningful scaling limit for $L\to\infty$
if positions $i_{k}$ and time $t$ are scaled diffusively according
to 
\begin{equation} \label{eq:scaling-limit}
i_{k} \to x_{k}=i_{k}/L\in[0,1],\quad t \to t/L^{2}.
\end{equation}
In this limit one defines the expectation value of on $n$-loops
\begin{equation} \label{eq:def_g_singular}
g_{t}^{s}(x_{1},\cdots,x_{n}):=\lim_{L\to\infty}L^{n-1}\mathbb{E}_{L^{2}t}[G_{i_{1}i_{2}}G_{i_{2}i_{3}}\cdots G_{i_{n}i_{1}}],
\end{equation}
which satisfies
\begin{align}\label{eq:g_n^s_evolution}
&(\partial_{t}-\Delta)g_{t}^{s}(x_{1},\cdots,x_{n})\\
&=\sum_{i<j}^{n}2\partial_{i}\partial_{j}
\left(\delta(x_{i},x_{j}) g_{t}^{s}(x_{i},\cdots,x_{j-1})g_{t}^{s}(x_{j},\cdots,x_{i-1})\right) \nonumber
\end{align}
where $\Delta\equiv\sum_{i=1}^{n}\Delta_{x_{i}}$, $\partial_{i}\equiv\partial_{x_{i}}$ and $\delta(x_i,x_j)=\delta(x_i-x_j)$. Pictorially, this equation reduces the evolution of an $n$-loop expectation value to diffusion sourced by the product of two loop expectation values with less nodes that emerge by pinching the original loop along the nodes $x_i$ and $x_j$,
\begin{equation}
\raisebox{-0.5\height}{\includegraphics{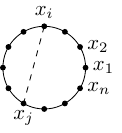}}
\longrightarrow
\raisebox{-0.5\height}{\includegraphics{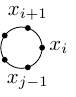}}
\,
\raisebox{-0.5\height}{\includegraphics{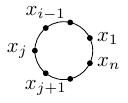}}.
\end{equation}
We therefore say that the evolution for $g^s_t$ has a \textit{triangular structure}.

In section \ref{subsec:open_boundary_scaling_limit} we will show that this equation is also valid for the open Q-SSEP, if we require $g^s_t(x_1,\cdots,x_n)$ on the boundary $x_i\in\partial[0,1]$ to be equal to its steady state value at all times. Due to the triangular structure of the equation this behaviour can be traced back to the fermion density, i.e.\ the $1$-loop $g^s_t(x)$, which approaches its steady state value on the boundary immediately, in contrast to its evolution in the bulk which happens on a much slower time scale. This is explained in section \ref{subsec:Bulk-vs-Boundary}.

As discussed below \eqref{eq:factorisation_of_loops}, loop expectation values jump by an order of $L$ if two indices become equal. This is reflected by the fact that solutions $g^s_t(x_1,\cdots,x_n)$ of \eqref{eq:g_n^s_evolution} become singular whenever two of its arguments $x_i=x_j$ are equal, hence the choice for the superscript "s". We show this in appendix \ref{app:Singular-behavior-of} for the example of $n=2$.

In appendix \ref{app:derivation-connected-correlations} we derive the time evolution of connected loop expectation values in the scaling limit. Connected $n$-loop expectation values are defined as
\begin{equation} \label{eq:def_g}
g_{t}(x_{1},\cdots,x_{n}):=\lim_{L\to\infty}L^{n-1}\mathbb{E}_{L^{2}t}[G_{i_{1}i_{2}}G_{i_{2}i_{3}}\cdots G_{i_{n}i_{1}}]^c.
\end{equation}
We find that they satisfy
\begin{align}\label{eq:g_n_evolution}
&(\partial_{t}-\Delta)g_{t}(x_{1},\cdots,x_{n})
\\ \nonumber
&=\sum_{i<j}^{n}2\,\delta(x_{i},x_{j})\partial_{i}g_{t}(x_{i},\cdots,x_{j-1})\partial_{j}g_{t}(x_{j},\cdots,x_{i-1}),
\end{align}
with boundary conditions
\begin{equation}
g_t(x_{1},\cdots,x_{n})=
\begin{cases}
n_a, \,n_b \text{ for } n=1 \text{ and } x=0, \,1 \\
0  \text{ for } n\ge 2 \text{ and some } x_i\in\{0,1\}, 
\end{cases}
\end{equation}
where only the one-point function  (fermion density) depends on particle density of the two reservoirs $n_a$ and $n_b$.
This equation produces indeed solutions that are continuous if two arguments become equal, see appendix \ref{app:Singular-behavior-of}. The equation differs from \eqref{eq:g_n^s_evolution} by the relative position of the derivatives and the delta function.

The derivation of equation \eqref{eq:g_n_evolution} crucially depends on the fact that loop expectation values (moments) and their connected parts (cumulants) are related by a sum over non-crossing partitions such as in \eqref{eq:moment-cumulant_Q-SSEP}. In the scaling limit this relation becomes
\begin{equation}\label{eq:g^s_expansion_in_g}
g_{t}^{s}(x_1,\cdots,x_n)=\sum_{\pi\in NC(n)}\delta_\pi\prod_{b\in\pi^*} g_t(x_{b(1)},\cdots,x_{b(|b|)}),
\end{equation}
where 
\begin{equation}
\delta_\pi \equiv
\prod_{b\in\pi^*} \delta(x_{b(1)},\cdots,x_{b(|b|)})
\end{equation}
is defined in analogy to \eqref{eq:kronecker-delta_pi}, but for continuous variables.

Finally, in section \ref{subsec:steady_state_solution} we show how to construct a very simple steady state solution for connected loop expectation values. This exploits the insight below \eqref{eq:moments_G_expansion} where we noted that the presence of the $\delta_\pi$ prevents us from viewing connected loop expectation values as the free cumulants of the measure $\mathbb E_t$. If we remove the $\delta_\pi$, then connected loop expectation values are, by definition, the free cumulants of a new measure $\varphi_t$ that is necessarily different from $\mathbb E_t$. Surprisingly $\varphi_t$ has a very simple steady state solution from which the connected loop expectation values $g_\infty$ can be determined recursively.

\subsection{Introduction to the model}\label{subsec:intro_Q-SSEP}

The quantum symmetric simple exclusion process was first introduced
in \cite{Bauer2017Stochastic} (closed case) and \cite{Bernard2019Open}
(open case) and was further elaborated in \cite{Bauer2019Equilibrium,Bernard2022DynamicsClosed,Bernard2021Solution}.

\paragraph{Definition.}

The model describes a one-dimensional chain with $L$ sites on which
spinless fermions hop to neighbouring sites with random amplitudes. The boundary sites are coupled to particle reservoirs (comment \footnote{The model can be defined on any graph but we are dealing with Q-SSEP on a line.}). The bulk
evolution of the system is stochastic and unitary. We describe it in
terms of the systems density matrix $\rho_t$ as
\begin{equation}
\rho_{t}\rightarrow \rho_{t+dt} = e^{-idH_{t}}\rho_{t}e^{idH_{t}},\label{eq:Q-SSEP_evolution}
\end{equation}
 where the Hamiltonian increment is defined as
\[
dH_{t}=\sum_{j=1}^{L-1}c_{j+1}^{\dagger}c_{j}dW_{t}^{j}+c_{j}^{\dagger}c_{j+1}d\overline{W}_{t}^{j},
\]
with $c_{j}^{\dagger}$ ($c_{j}$) fermion creation (annihilation)
operators at site $j$ with the usual commutation relations $\{c_{j}^{\dagger},c_{k}\}=\delta_{jk}$
and $\{c_{j}^{\dagger},c_{k}^{\dagger}\}=\{c_{j},c_{k}\}=0$. The noisy Hamiltonian increment depends on a complex Brownian motion $dW_{t}^{j}$ which determines
the random hopping amplitude along the edge $(j,j+1)$ at time $t$. There is one complex Brownian motion per edge.
This means that $W_{t}^{j}$ at each instance $t$ is a centred Gaussian
random variable that has independent increments
$dW_{t}^{j}=W_{t+dt}^{j}-W_{t}^{j}$ with variance $\mathbb{E}[dW_{t}^{j}d\overline{W}_{t'}^{k}]=J\delta^{j,k}dt$
if $t=t'$, and zero otherwise. $J$ is a rate parameter with units
one over time and we set $J=1$ in the following. 

To the stochastic and unitary bulk evolution (\ref{eq:Q-SSEP_evolution}),
we add a deterministic but dissipative Lindbladian evolution 
\[
\mathcal{L}_\mathrm{bdry}=\alpha_{1}\mathcal{L}_{1}^{+}+\beta_{1}\mathcal{L}_{1}^{-}+\alpha_{L}\mathcal{L}_{L}^{+}+\beta_{L}\mathcal{L}_{L}^{-},
\]
 that represents the interaction with the reservoirs. The operator
$\mathcal{L}_{j}^{+}(\bullet)=c_{j}^{\dagger}\bullet c_{j}-\frac{1}{2}\{c_{j}c_{j}^{\dagger},\bullet\}$
models particle injection and is multiplied by the injection rate
$\alpha_{j}$ while $\mathcal{L}_{j}^{-}(\bullet)=c_{j}\bullet c_{j}^{\dagger}-\frac{1}{2}\{c_{j}^{\dagger}c_{j},\bullet\}$
models particle extraction with rate $\beta_{j}$. For example, the density matrix of an
isolated empty site $\tau{}_{t}=|0\rangle\langle0|$ that evolves
according to $\partial_{t}\tau_{t}=\alpha\mathcal{L}^{+}(\tau_{t})$
will be occupied after a time interval $dt$ with probability $\alpha dt$. That is,
\[
|0\rangle\langle0|\to\alpha dt|1\rangle\langle1|+(1-\alpha dt)|0\rangle\langle0|.
\]

The full evolution of the systems density matrix $\rho_t$ can be expressed as a stochastic differential equation (SDE) in Itô convention (with Itô rules $dW_{t}^{j}d\overline{W}_{t}^{k}=\delta^{j,k}dt$)
by expanding (\ref{eq:Q-SSEP_evolution}) up to order $dt$,
\begin{equation}
d\rho_{t}=-i[dH_{t},\rho_{t}]-\frac{1}{2}[dH_{t},[dH_{t},\rho_{t}]]+\mathcal{L}_\mathrm{bdry}(\rho_{t})dt.\label{eq:rho_Q-SSEP_evolution}
\end{equation}
Occasionally we will refer to the ``closed case'', by which we mean
that there are no boundary reservoirs and the one-dimensional
chain is closed periodically
such that sites $1\equiv L$ are identified.

\paragraph{Relation to the classical SSEP.}

The name of this model is inherited from the classical symmetric simple exclusion process (SSEP). The latter can be obtained from Q-SSEP in the special case where one is just interested in the mean density matrix $\bar{\rho}_{t}:=\mathbb{E}[\rho_{t}]$, where the expectation $\mathbb{E}[\cdots]$ is taken with respect to the different realisations of the Brownian motions. This matrix evolves according to a Lindblad equation,
\begin{align*}
\partial_{t}\bar{\rho_{t}} =\mathcal{L}(\bar{\rho_{t}})
= \sum_{j=1}^{L-1}\mathcal{L}^\mathrm{edge}_j(\bar{\rho_{t}}) +\mathcal{L}_\mathrm{bdry}(\bar{\rho_{t}}),
\end{align*}
with $\mathcal{L}^\mathrm{edge}_j(\bar{\rho})=l_{j}^{-}\bar{\rho}l_{j}^{+}+l_{j}^{+}\bar{\rho}l_{j}^{-}-\frac{1}{2}\{l_{j}^{+}l_{j}^{-}+l_{j}^{-}l_{j}^{+},\bar{\rho}\} $
where $l_{j}^{+}=c_{j+1}^{\dagger}c_{j}$ and $l_{j}^{-}=c_{j}^{\dagger}c_{j+1}$.
Writing $\bar{\rho}_t$ in the occupation number basis, this Lindbladian
evolution preserves its diagonal structure, while off-diagonal elements
vanish exponentially in time. The diagonal elements of $\bar{\rho}_t$
provide the probability that the system is in one of the $2^{L}$
states with well defined occupation number $\hat{n}_{i}=c_{i}^{\dagger}c_{i}$
on each site. This corresponds to a configuration of the classical
SSEP, where one specifies the number of particles $n_{i}=0,1$ on
each site. One can see easily that the Lindbladian evolution of the
diagonal elements of $\bar{\rho}_{t}$ precisely corresponds to the
Markov process of SSEP: During a time interval $dt$ a particle in
the bulk can jump to the left or right neighbouring site with probability
$dt$ if the site is empty and particles get
injected (extracted) at the boundaries with probability $\alpha_{i}dt$ ($\beta_{i}dt$).
This correspondence can also be formulated via the moment generating
function,
\begin{equation}
\langle e^{\sum_{i}a_{i}n_{i}}\rangle_\mathrm{ssep}=\mathrm{Tr}\big(\bar{\rho}\,  e^{\sum_{i}a_{i}\hat{n}_{i}}\big),\label{eq:seep-Q-SSEP}
\end{equation}
and it illustrates the well known correspondence between Markov processes and Lindbladian evolutions.
The relation between Q-SSEP and free cumulants we shall describe below implies a new relation between the classical SSEP and free probability \cite{Bauer2022Bernoulli}.

\paragraph{Fluctuating coherences.}

If we go beyond the mean dynamics, then Q-SSEP has an additional structure,
which describes purely quantum mechanical effects such as entanglement. This structure is inscribed into the coherences $G_{ij}(t)=\mathrm{Tr}(\rho_{t}c_{i}^{\dagger}c_{j})$. While the coherences vanish in mean exponentially with time (we
saw that the mean corresponds to the classical SSEP and there are
no quantum correlations left), their fluctuations survive the long time limit, 
although they are sub-leading in the system size \cite{Bernard2019Open}.
For example, at large system size $L\to\infty$ with $x=i/L\le y=j/L$
fixed, the connected quadratic fluctuation (or 2nd cumulant) in the
steady state is
\[
\mathbb{E_{\infty}}[G_{ij}G_{ji}]^{c}\sim\frac{1}{L}(\Delta n)^{2}x(1-y),
\]
where $\Delta n=n_{a}-n_{b}$ is the difference in the particle density
between the boundary reservoirs, see (\ref{eq:reservoir_denisty}). Here we adopted again the point of view where the time dependence of $G_{ij}(t)$ is transfered to the measure $\mathbb E_t$ and the steady measure is denoted by $\mathbb E_\infty$.

Let us also note that since the Hamiltonian is quadratic in fermionic creation and annihilation operators, the coherences
$G_{ij}$ completely characterise the state of the system due to Wick's
theorem. From (\ref{eq:rho_Q-SSEP_evolution}) one finds that their
time evolution is given by the SDE 
\begin{widetext}
\begin{align} 
dG_{ij} =&-i(G_{i,j-1}dW_{t}^{j-1}+G_{i,j+1}d\overline{W}_{t}^{j}-G_{i-1,j}d\overline{W}_{t}^{i-1}-G_{i+1,j}dW^{i}) +\delta_{ij}(G_{i+1,i+1}+G_{i-1,i-1})dt  \label{eq:dGij} \\
 & -2G_{ij}dt -\sum_{p\in\{1,L\}}\left(\frac{1}{2}(\delta_{ip}+\delta_{jp})(\alpha_{p}+\beta_{p}-1)G_{ij}-\delta_{pi}\delta_{pj}\alpha_{p} \right)dt.\nonumber 
\end{align}
\end{widetext}

\paragraph{Three properties of $\mathbb E_t$.}
Firstly, the measure $\mathbb E_t$ of Q-SSEP possesses a local $U(1)$
invariance which can bee seen as follows: The multiplication with local phases, $G_{ij}\to\tilde{G}_{ij}=e^{-i\theta_{i}}G_{ij}e^{i\theta_{j}}$ leaves (\ref{eq:dGij}) invariant if also the Brownian motions are multiplied
by a phase $dW_{t}^{j}\to d\tilde{W}_{t}^{j}=e^{i(\theta_{j+1}-\theta_{j})}dW_{t}^{j}$.
Since $d\tilde{W}_{t}^{j}$ and $dW_{t}^{j}$ have the same distributions, also $G$ and $\tilde{G}$ have the same distribution if they agree at $t=0$, which means that initially $G$ is a diagonal matrix. In the scaling limit this is always the case since off-diagonal terms vanish exponentially fast in system size ($\sim e^{-L^{2}t}$ where $t$ is the rescaled time).

Secondly, $n$-loop expectation values in Q-SSEP scale as $L^{-n+1}$ if indices are distinct because this scaling gives rise to the non-trivial equation \eqref{eq:g_n^s_evolution}.

Thirdly, expectation values of products of loops factorise. This is shown in appendix G of \cite{Bernard2022DynamicsClosed} where the time evolution equation of loop expectation values \eqref{eq:g_n^s_evolution} is derived (for the closed case). The reason is that the time evolution equation for the leading order of a product of loop expectation values allows for factorised solutions. Since the initial condition is always factorised (the noise has not yet correlated the variables) the claim follows. 

This shows that for Q-SSEP properties \eqref{eq:u(1)-invariance}-\eqref{eq:factorisation_of_loops} are satisfied and as a result Q-SSEP has the expansion of loop expectation values into non-crossing partitions \eqref{eq:moment-cumulant_Q-SSEP} as claimed.

\subsection{Bulk vs.\ boundary thermalisation\label{subsec:Bulk-vs-Boundary}}

When we casually talk about bulk and boundary thermalisation, one
should keep in mind that the open Q-SSEP does not actually ``thermalise''.
At large times, the system approaches a steady state (comment \footnote{To be precise: if one is looking at the mean behaviour then the system approaches a mean steady state, but if one is looking at fluctuations the system is then approaching a steady distribution of states.}), 
in which observables do not depend on time any more longer. But they are not described by a thermal density matrix since there is a steady current, so that the system is out-of-equilibrium. 

Here we show that the density of fermions in the open
Q-SSEP approaches its steady state value much faster on the boundaries
than in the bulk. In particular, the decay times scale with $\mathcal{O}(L^2)$
in the bulk and $\mathcal{O}(1)$ on the boundary. Due
to the correspondence (\ref{eq:seep-Q-SSEP}) this property also hold
true for the classical SSEP (comment \footnote{Although we expect this result to be known, we did not find any paper on the classical SSEP where this is shown explicitly}).

Under the evolution outlined above the mean fermion density $n_{i}(t):=\mathbb{E}_t[\mathrm{Tr}(\rho c_{i}^{\dagger}c_{i})]$
evolves according to
\begin{align}
\partial_{t}n_{i}(t) & =\Delta n_{i}(t)+\sum_{p={1,L}}\delta_{ip}(\alpha_{p}-(\alpha_{p}+\beta_{p})n_{p}(t)),\label{eq:density_discrete}
\end{align}
where it is understood that the discrete Laplacian on the boundaries
is truncated, i.e. $\Delta n_{1}:=n_{2}-n_{1}$ and $\Delta n_{L}=n_{L-1}-n_{L}$.
Here we only discuss the special case, where the injection/extraction
parameters are such that $\alpha_{1}+\beta_{1}=1=\alpha_{L}+\beta_{L}$.
The general case is discussed in Appendix \ref{app:Derivation-density}
and works analogously with a little bit of help of numerical calculations.
In the special case, (\ref{eq:density_discrete}) can be solved analytically
\begin{equation}\label{eq:density_discrete_solution}
n_{j}(t)= \sum_{k=1}^{L}\sin(\frac{jk\pi}{L+1})\left(e^{(-2+2\cos(\frac{k\pi}{L+1}))t}(c_{k}-b_k)+b_k\right),
\end{equation}
where the coefficients $c_{k}$ are determined by the initial condition and $b_k:=\frac{2}{L+1}\frac{\alpha_{1}\sin(\frac{k\pi}{L+1})+\alpha_{L}\sin(\frac{Lk\pi}{L+1})}{2(1-\cos(\frac{k\pi}{L+1}))}$. This solution consists of two contributions:
A time dependent term that decays exponentially in time and a constant term 
that provides the steady state value and can be simplified to $n_{j}(\infty)=\frac{\alpha_{1}(L-j+1)+\alpha_{L}j}{L+1}$. 
We study the time scale with which the time dependent term decays on the
boundary compared to its decay in the bulk.

A bulk site is characterised by $j\sim aL$ with $a\sim\mathcal{O}(1)$.
Due to the factor $e^{(-2+2\cos(\frac{k\pi}{L+1}))t}\approx e^{-\pi^{2}k^{2}t/L^{2}}$
(for large $L$) the biggest contribution to the sum comes from the
term with $k=1$. Since we are in the bulk its amplitude is finite,
$\sin(j\pi/(L+1))\sim\sin(a\pi)$. Therefore we find a time scale
of $t_{\mathrm{decay}}\sim\mathcal{O}(L^{2})$,
\[
n_\mathrm{bulk}(t)\sim\mathrm{const.}e^{-\pi^{2}t/L^{2}}+n_\mathrm{bulk}(\infty).
\]

A site on the boundary is characterized by $j=1$ or $j=L$ and therefore
the amplitude of the $k=1$ term, $\sin(j\pi/(L+1))$, will be zero
for large $L$. A significant contribution therefore only comes from
terms where $k\sim bL$ with $b\sim\mathcal{O}(1)$, because then
the amplitudes $\sin(jk\pi/(L+1))\sim\sin(b\pi)$ or $\sim \sin(bL\pi)$
stay finite. The time scale with which these terms decay is of order
one, $t_{\mathrm{decay}}\sim\mathcal{O}(1)$,
\[
n_\mathrm{bdry}(t)\sim\mathrm{const.}\, e^{-b^{2}\pi^{2}t}+n_\mathrm{bdry}(\infty).
\]

Note that the value of the density on the boundaries after a time
of $\mathcal{O}(1)$ is that of the steady state. For general injection/extraction
parameters these values are as in the classical SSEP,
\begin{align}
n_{1}(\infty) =n_{a}:=\frac{\alpha_{1}}{\alpha_{1}+\beta_{1}} &,\  n_{L}(\infty)=n_{b}:=\frac{\alpha_{L}}{\alpha_{L}+\beta_{L}}.\label{eq:reservoir_denisty}
\end{align}
We conclude that time scales with which the
boundary and bulk approach their steady state values are separated by an order of $L^2$. In the scaling
limit (\ref{eq:scaling-limit}) where $t\to t/L^2$ denotes the rescaled time, this means 
that boundaries are equal to the steady state values at all times, while
in the bulk it takes $t_\mathrm{decay}\sim\mathcal{O}(1)$ of time.

\subsection{Open boundary scaling limit\label{subsec:open_boundary_scaling_limit}}

The basic question is whether the scaling limit \eqref{eq:scaling-limit} introduced for the closed Q-SSEP is meaningful also in the open case. In other words, does $g^s_t$ satisfy a non-trivial equation in this limit? 
\paragraph{Density}
We first answer this question on the level of the density $n_i(t)=\mathbb E_t[G_{ii}]$. It satisfies $L$  coupled differential equations in time given by \eqref{eq:density_discrete}. The aim is to represent these by a single partial differential equation in space and time for a continuous density $\rho(x,t)\approx n_{Lx}(L^{2}t)$. Since it is easily seen that the bulk satisfies a pure diffusion equation the remaining question is what are the correct boundary conditions. The answer is provided by the observation made in the last section. In the scaling limit, the boundaries immediately take their steady state values. Therefore,
\begin{align}
\partial_{t}\rho(x,t)=\partial_{x}^{2}\rho(x,t),\label{eq:density_scaling}\\
\rho(0,t)=n_{a},\, \rho(1,t)=n_{b}.
\end{align}

\begin{figure}[h]
\vskip 0.5 truecm
\centering\includegraphics[width=0.4\textwidth]{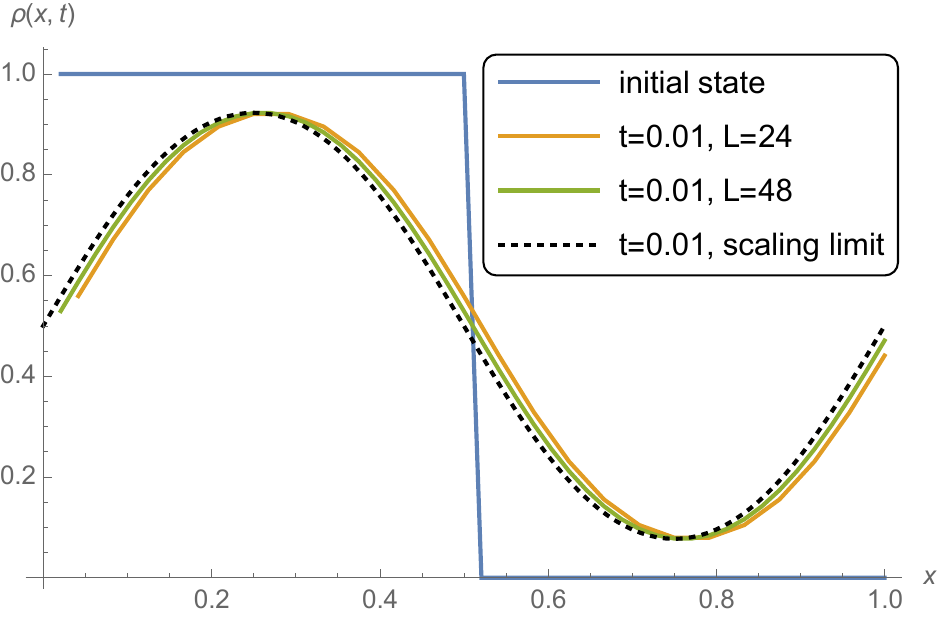}\caption{The discrete fermion density $n_{Lx}(L^{2}t)$ for system sizes $L=24$
and $L=48$ together with the scaling limit $\rho(x,t)$ at $t=0.01$
as a function of space $x=i/L$. The extraction and injection rates
are $\alpha_{1}=\beta_{1}=\alpha_{L}=\beta_{L}=1$ and don't fit the
initial conditions. The agreement is very good.\label{fig:discrete_continuous_density}}
\vskip 0.5 truecm
\end{figure}

This is confirmed numerically in Figure (\ref{fig:discrete_continuous_density}),
where a solution for the discrete density $n_{i}(t)$ is compared
to the solution for the continuous density $\rho(x,t)$. In particular
the boundary values agree. 

\paragraph{Higher order fluctuations.}
The claim we make here is that the exact same equation \eqref{eq:g_n^s_evolution} that holds in the closed case also holds in the open case if we subject $g^{s}$
to boundary conditions that are equal to its steady state value.
These are known from \cite{Bernard2019Open} for its connected part:
$g_{\infty}(x_{1},\cdots,x_{n})=0$ for all $n\ge 2$ and $(x_{1},\cdots,x_{n})\in\partial[0,1]^{n}$. 

We test this claim on the level of the $2$-loop expectation values, for which \eqref{eq:g_n^s_evolution} simplifies to
\begin{equation}
(\partial_{t}-\Delta)g_{t}^{s}(x,y)=2\partial_{x}\partial_{y}(\delta(x-y)g_{t}(x)g_{t}(y)),\label{eq:g_2^s_evolution}
\end{equation}
and we identified $1$-loops with the density, $g_{t}(x)=\rho(x,t)$.
From this it follows that the time evolution of the connected correlation functions with
appropriate boundary conditions is
\begin{align}
(\partial_{t}-\Delta)g_{t}(x,y) & =2\delta(x-y)\partial_{x}g_{t}(x)\partial_{y}g_{t}(y),\label{eq:g_2_evolution}\\
g_{t}(x,y) & =0\text{ if }(x,y)\in\partial[0,1]^{2},\nonumber 
\end{align}
which is a again a diffusion equation, but with a new source term.
To see this, note that the connected correlation function is defined by
\begin{align*}
g_{t}(x,y)= & \lim_{L\to\infty}L(\mathbb{E}_{L^{2}t}[G_{ij}G_{ji}]-\delta_{ij}\mathbb{E}_{L^{2}t}[G_{ii}]\mathbb{E}_{L^{2}t}[G_{jj}])\\
= &\ g_{t}^{s}(x,y)-\delta(x-y)g_{t}(x)g_{t}(y).
\end{align*}

\begin{figure}[h]
\vskip 0.5 truecm
\centering\includegraphics[width=0.4\textwidth]{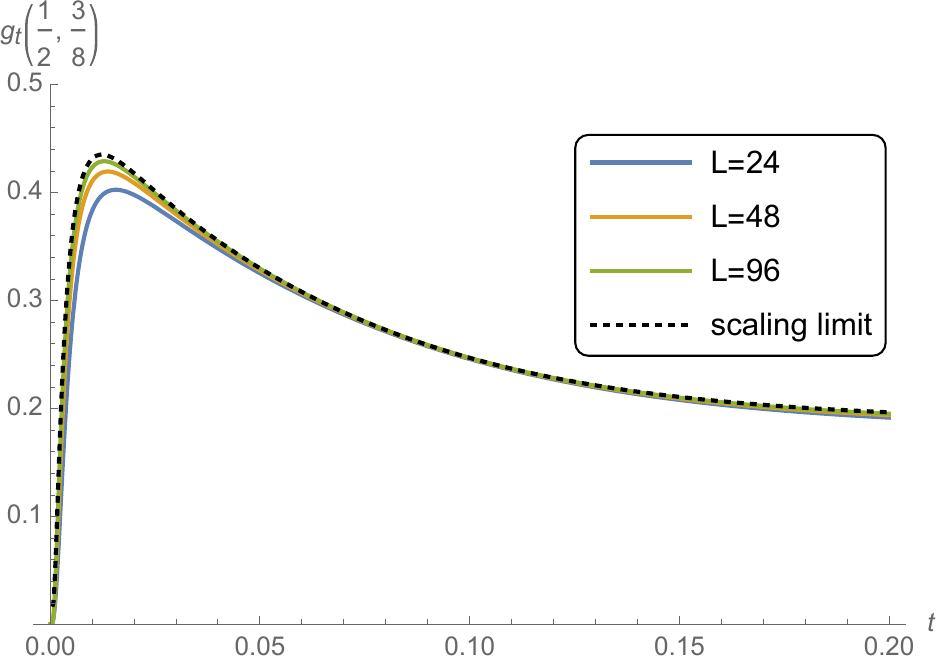}\caption{Boundary conditions that fit the initial
domain wall state ($n_{a}=1$, $n_{b}=0$).}
\label{fig:g_2 with fitting initial}
\vskip 0.5 truecm
\end{figure}

\begin{figure}[h]
\vskip 0.5 truecm
\centering\includegraphics[width=0.4\textwidth]{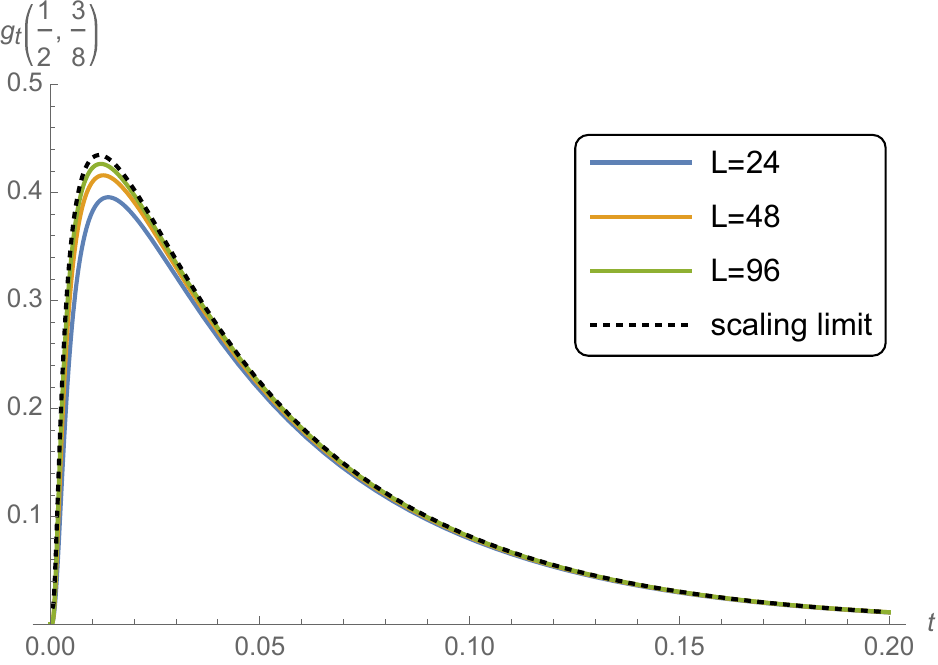}\caption{Boundary conditions that do not fit
the initial domain wall state ($n_{a}=1/2$, $n_{b}=1/2$).}
\label{fig:g_2 with unfitting initial}
\vskip 0.5 truecm
\end{figure}

The analytic solution of \eqref{eq:g_2_evolution} (which is constructed in appendix
\ref{app:Analytic-solution-of}) is compared to a numerical solution
of the discrete evolution equations of $L(\mathbb E_{L^{2}t}[G_{ij}G_{ji}]-\delta_{ij}\mathbb E_{L^{2}t}[G_{ii}]^2)$
for given $L$ -- which can be derived from (\ref{eq:dGij}).
Figures \ref{fig:g_2 with unfitting initial} and \ref{fig:g_2 with fitting initial}
show the result of this comparison for different system sizes $L$.
The agreement is excellent.

Notice that the injection and extraction rates enter (\ref{eq:g_2_evolution}) only through the source term, since the density $g_{t}(x)$
depends on these rates through its own boundary conditions. Apart
from that, the equation for the connected $2$-loop expectation values $g_{t}(x,y)$ (and for
all higher order loops) never explicitly depend on the injection
and extraction rates. As a consequence, without loss of generality, 
we use the convention
\[ n_a=0,\quad n_b=1,\]
in the following, through out the paper. 
In particular, the density in the steady state $g_\infty(x)=n_a + x(n_b-n_a)$ reduces to $g_\infty(x)=x$ in this convention.

As a result, one can now derive the time evolution equation of connected loop expectation values in the scaling limit with open boundaries (see appendix \ref{app:derivation-connected-correlations}). One finds,
\begin{align}
&(\partial_{t}-\Delta)g_{t}(x_{1},\cdots,x_{n})
\\ \nonumber
&=\sum_{i<j}^{n}2\,\delta(x_{i},x_{j})\partial_{i}g_{t}(x_{i},\cdots,x_{j-1})\partial_{j}g_{t}(x_{j},\cdots,x_{i-1}),
\end{align}
with boundary conditions
\begin{equation}
g_t(x_{1},\cdots,x_{n})=
\begin{cases}
n_a, \,n_b \text{ for } n=1 \text{ and } x=0, \,1 \\
0  \text{ for } n\ge 2 \text{ and some } x_i\in\{0,1\}.
\end{cases}
\end{equation}

\subsection{Free cumulants in Q-SSEP}\label{subsec:free_cumulants_in_qssep}
Comparing the moment-cumulant relation \eqref{eq:moment-cumulant_Q-SSEP} (valid for any system satisfying \eqref{eq:u(1)-invariance}-\eqref{eq:factorisation_of_loops}, in particular for Q-SSEP) to the definition of free cumulants \eqref{eq:def_free_cumulants} one would be tempted to identify 
\begin{equation}\label{eq:identification}
\mathbb E_t[G_{i_1i_2}\cdots G_{i_ni_1}]^c\sim\kappa_{n}(G_{i_1i_2},\cdots,G_{i_ni_1})
\end{equation}
with $\kappa_n$ the free cumulants of $\mathbb E_t$. However, this is not correct due to the presence of $\delta_\pi$. If we nontheless insist on the identification \eqref{eq:identification} then the connected loop expectation values $\mathbb E[G_{i_1i2}\cdots G_{i_ni_1}]^c$ are, by definition, the free cumulants of a new measure $\varphi_t$ that is different from $\mathbb E_t$. That is,
\begin{equation}\label{eq:phi_discrete}
\begin{aligned}
&\mathbb \varphi_t (G_{i_1i_2}\cdots G_{i_ni_1})
\\
&:=\sum_{\pi\in NC(n)}\prod_{b\in \pi}\mathbb E_t[G_{i_{b(1)}i_{b(2)}}\cdots G_{i_{b(|b|)}i_{b(1)}}]^c.
\end{aligned}
\end{equation}
In terms of $g_{t}(x_{1},\cdots,x_{n})\sim L^{n-1}\mathbb{E}_{L^{2}t}[G_{i_{1}i_{2}}\cdots G_{i_{n}i_{1}}]^c$ the new measure $\varphi_t$ defines a function (which we denote by the same name)
of continuous positions and rescaled time 
\begin{equation}\label{eq:phi_g}
\varphi_{t}(x_1,\cdots,x_n)=\sum_{\pi\in NC(n)}\prod_{b\in\pi} g_t(x_{b(1)},\cdots,x_{b(|b|)}).
\end{equation}
The time evolution of $\varphi_t$ is found to satisfy exactly the same equation as $g_t$ (see appendix \ref{app:derivation-new-measure})
\begin{align}\label{eq:phi_equation}
&(\partial_{t}-\Delta)\varphi_{t}(x_{1},\cdots,x_{n})
\\ \nonumber
&=\sum_{i<j}^{n}2\,\delta(x_{i},x_{j})\partial_{i}\varphi_{t}(x_{i},\cdots,x_{j-1})\partial_{j}\varphi_{t}(x_{j},\cdots,x_{i-1}).
\end{align}
However the boundary conditions are different. If some $x_i\in\{0,1\}$ lies on the boundary, then 
\begin{align}\label{eq:phi_boundary}
\varphi_{t}(x_{1},\cdots,x_i,\cdots,x_{n})= x_i\, \varphi_{t}(x_{1},\cdots,\hat{x}_{i},\cdots,x_{n}) 
\end{align} 
where the hat on $\hat{x}_{i}$ indicates that $x_{i}$ is missing from the
set $\{x_{1},\cdots,x_{n}\}$.

\subsection{Steady state solution - inspired by free probability}\label{subsec:steady_state_solution}
The striking observation by Biane \cite{Biane2021Combinatorics} was that $\varphi_t$, defined as a sum over non-crossing partitions of products of connected loop expectation values, has a very simple solution in the steady state. In appendix \ref{app:steady_measure} we show that for $t\to\infty$, our equation for $\varphi_t$ is indeed solved by 
\begin{equation}
\varphi_{\infty}(x_{1},\cdots,x_{n})=\min(x_{1},\cdots,x_{n})
\end{equation}

As a consequence, we can use \eqref{eq:phi_g} to recursively
reconstruct the connected correlations functions $g_{\infty}$ in the steady
state. This works similarly to \eqref{eq:moments_free_cumulants_4}
and \eqref{eq:free_cumulants_moments_4}. Denoting $\min(x_{1},\cdots,x_{n})=:x_{1}\wedge\cdots\wedge x_{n}$,
we have
\begin{align*}
g_{\infty}(x_{1})= &\ x_{1}\\
g_{\infty}(x_{1},x_{2})= &\ x_{1}\wedge x_{2}-x_{1}x_{2}\\
g_{\infty}(x_{1},x_{2},x_{3})= &\ x_{1}\wedge x_{2}\wedge x_{3}-x_{1}(x_{2}\wedge x_{3})_{\circlearrowleft3}+2x_{1}x_{2}x_{3},
\end{align*}
while the four point function $g_{\infty}(x_{1},x_{2},x_{3},x_{4})$ reads
\begin{align*}
&\ x_{1}\wedge x_{2}\wedge x_{3}\wedge x_{4}-x_{1}(x_{2}\wedge x_{3}\wedge x_{4})_{\circlearrowleft4}
\\
-(&x_{1}\wedge x_{2})(x_{3}\wedge x_{4})_{\circlearrowleft2}+2x_{1}x_{2}(x_{3}\wedge x_{4})_{\circlearrowleft4}
\\
+&x_{1}x_{3}(x_{2}\wedge x_{4})_{\circlearrowleft2}-5x_{1}x_{2}x_{3}x_{4},
\end{align*}
where $\cdots_{\circlearrowleft q}$ denotes the sum of all
terms obtained by $q$ successive cyclic permutations of the
arguments of the term in question. Note that due to the absence of crossing partitions, $g_\infty(x_1,\cdots,x_n)$ is no longer invariant under the permutation of its arguments for $n\ge4$. In the example above it is the term $(x_{1}\wedge x_{2})(x_{3}\wedge x_{4})_{\circlearrowleft2}$ which prevents $g_\infty(x_1,\cdots,x_4)$ from being invariant under the exchange of $x_2$ and $x_3$. 

Indeed, ordering the variables as $0\le x_1\le x_2\le x_3\le x_4\le1$, we get $g_{\infty}(x_{1},x_{2})=x_1(1-x_2)$ and $g_{\infty}(x_{1},x_{2},x_{3})=x_1(1-2x_2)(1-x_3)$ and
\begin{align*}
   g_{\infty}(x_{1},x_{2},x_{3},x_4) = &\ x_1(1- 3 x_2-2x_3+5x_2x_3)(1-x_4) ~, \\
     g_{\infty}(x_{1},x_{3},x_{2},x_4) = &\ x_1(1- 4x_2-x_3+5x_2x_3)(1-x_4) ~,
\end{align*}
whereas $g_{\infty}(x_{1},x_{3},x_{4},x_2)=g_{\infty}(x_{1},x_{2},x_{3},x_4)$, in agreement with \cite{Bernard2019Open}.

The derivation of the steady state solution presented here provides an alternative proof of Biane's formula
(6.1) in \cite{Biane2021Combinatorics} that relates the steady state
connected correlations of the open Q-SSEP to free cumulants of the
measure $\varphi_{\infty}$. Our derivation extends this relation to finite times, though an explicit solution for $\varphi_t$ at finite times seems out of reach.

Note that the measure $\varphi_\infty$ can be realized as the Lebesgue
measure on the interval $[0,1]$ of the indicator function $\mathbb{I}_{x}:=1_{[0,x]}$,
e.g.
\begin{equation}
\varphi_\infty(x,y):=\int_{0}^{1}\mathbb I_x(z)\mathbb I_y(z)dz=\min(x,y)
\end{equation}
It is surprising that $\varphi_\infty$ has a realization in terms of commuting variables,
since free cumulants usually appear in a setting of non-commuting
variables. At finite times (\ref{eq:phi_equation}) suggests that $\varphi_{t}$ is not invariant under a permutation of its arguments.

\section{Conclusion }\label{sec:conclusion}
In this paper we presented two main points. (1) A general argument why the fluctuations of spatial coherences in one dimensional mesoscopic quantum systems could be well described by the framework of free probability theory. (2) A precise calculation that shows that the model Q-SSEP has a mathematical structure that fits into the framework of free probability and we used this structure to derive the time evolution of connected correlation functions. Specific to the open Q-SSEP is the observation that the density approaches its steady state value much faster on the boundary than in the bulk which we used to formulate the correct boundary conditions for the time evolution of coherences in the scaling limit.

In both cases the link to free probability can be reduced to three properties of the noise expectation value: (i) local $U(1)$ invariance, (ii) a large deviation scaling of correlation functions $\mathbb{E}[G_{i_{1}i_{2}}\cdots G_{i_{n}i_{1}}]\sim L^{-n+1}$, and (iii) the fact that expectation values of products of loops factorise. It is not surprising that the same three properties are responsible for the fact that a relation with free probability has been observed in the context of the eigenstate thermalisation hypothesis (ETH).

We should also stress that the link with free probability in the context of coherent fluctuations in mesoscopic systems is in fact not particular to the systems being out-of-equilibrium. Rather, this link emerges from a coarse-grained description under the assumption that, locally in space (i.e.\ within ballistic cells) and on time scales much shorter than the diffusion time, the system is ergodic. Such an assumption allows to introduce the noise average as an average over all possible unitary transformation that the system could have undergone locally (i.e.\ within ballistic cells) and this noise average satisfies the three properties above. 

However, we admit that the general picture developed in section \ref{sec:general_picture} is certainly oversimplified and calls for more details. Firstly, the argument assumes a separation of time scales which gives rise to a crossover from ballistic to diffusive transport at some length scale $\ell$. One should provide criteria on the Hamiltonian for when this separation of time scales is satisfied. Secondly, one should explore the physical meaning of the length $\ell$ of ballistic cells. In the introduction we crudely argued that $\ell$ could be related to the mean free path of electrons in a disordered metal. And thirdly, a better understanding of the perturbative expansion of an interacting Hamiltonian -- the argument needed to derive the scaling of loop expectation values from classical MFT -- will allow us to characterise the domain of validity of this theory, that is for which class of systems it should be applicable.

To test the validity of this general picture, we plan to conduct numerical tests on more physical models. A first candidate would be a Floquet Heisenberg XXZ model with a staggered magnetic field, which breaks integrability but conserves the local $U(1)$-invariance. The time-dependence of the Hamiltonian ensures that energy is not conserved which would otherwise represent another conserved quantity.

Such numerical studies would also answer the question if Q-SSEP describes coherent fluctuations in a larger class of systems through the identification of sites in Q-SSEP with ballistic cells -- an idea we developed in the introduction.  This interpretation is supported by a recent work \cite{Jin2022Exact} where the authors use a decomposition into equilibrated and statistically independent cells to characterise transport properties of quantum stochastic Hamiltonians. In particular for the so-called \textit{dephasing model}, i.e.\ free fermions with an independent Brownian noise on each site, they find that the size of the cell scales as $1/\gamma$ where $\gamma$ is the strength of the noise. This result is interesting, since the Q-SSEP can be obtained as a limit of the dephasing model for strong noise $\gamma\to\infty$ \cite{Bauer2017Stochastic}. In this limit, the size of the cell becomes zero which corresponds to the idea that for Q-SSEP the mean free path $\ell$ has effectively been shrunk to the lattice spacing $a_\mathrm{uv}$, i.e.\ the ballistic cell contracts to a single site.

In spite of the progress reported here we believe that it remains a challenge to construct a {\it quantum mesoscopic fluctuation theory} describing fluctuations of quantum coherences in generic diffusive many-body systems at coarse-grained mesoscopic scales.

\bigskip

{\bf Acknowledgements.} 
This paper has been submitted simultaneously with "Eigenstate Thermalization Hypothesis and Free Probability"  by S. Pappalardi, L. Foini and J. Kurchan \cite{Pappalardi2022ETH}, where the relation between these two frameworks is discussed.  The occurrence of free probability  in both problems has a similar origin:  the coarse-graining at microscopic either spatial or energy scales, and the unitary invariance at these microscopic scales.  Thus the use of free probability tools promises to be ubiquitous in chaotic or noisy many-body quantum systems.
We thank Fabian Essler and Adam Nahum for numerous discussions on this topic. LH thanks Tony Jin for discussions about the general picture of fluctuating mesoscopic systems. This work was in part supported by CNRS, by the ENS and by the ANR project “ESQuisses”, contract number ANR-20-CE47-0014-01.


\appendix

\begin{widetext} 

\section{Measuring coherences\label{app:Measuring-Coherences}}

To gain a better understanding of the coherences $G_{ij}=\mathrm{Tr}(\rho c_{j}^{\dagger}c_{i})$
we outline an experiment to measure them that was proposed in \cite{Gullans2019Entanglement}.
The setup is shown in Figure (\ref{fig:experiment}). The idea is
to probe the system at spatially separated places and let the two
signals interfere before each output is measured separately. Let us
outline the steps of the measurement protocols in detail.

\begin{figure}[h]
\includegraphics{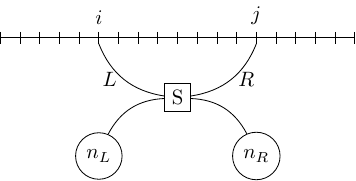}
\hspace{1cm}(a)
\includegraphics{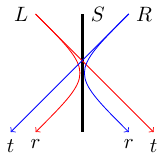}
\hspace{1cm}(b)
\includegraphics{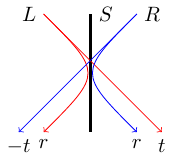}
\caption{Two wires are attached to the system at sites $i$ and $j$ such that
only one fermion can enter at a time. First the fermions in the wire
are allowed to interact via the beam splitter $S$. Then their occupation
number $n_{L}$ and $n_{R}$ is measured on each side. In the first
measurement (a) one uses a symmetric beam splitter, which allows to
measure the imaginary part of $G_{ij}$. In the second measurement
(b), one needs to use a beam splitter where the fermion that is transmitted
from $R$ to $L$ accumulates a phase $\pi$, while it does not accumulate
this phase when being transmitted in the other direction. In this
way one can measure the real part of $G_{ij}$. \label{fig:experiment}}
\end{figure}

\begin{itemize}
\item The total state of system, left and right wire is described by a state
in the Hilbert space $\mathcal{H}_{S}\otimes\mathcal{H}_{L}\otimes\mathcal{H}_{R}$.
Let us assume that the system is in a pure state and that initially
the wires are empty and not yet coupled to the system,
\[
|\psi^{(0)}\rangle=|\psi_{S}\rangle|0,0\rangle.
\]
\item Now we couple the two wires to the system. A very simple description
of this coupling could be given by the unitary evolution with $U_{int}=e^{-i\lambda(c_{L}^{\dagger}c_{i}+c_{R}^{\dagger}c_{j}+h.c.)}$,
where $\lambda$ is the product of coupling strength and the time
during which we allow the wires to couple to the system, and $c_{L}$
($c_{R}$) are fermionic operators on the left (right) wire. If we
tune the coupling strength and duration such that $\lambda\ll1$ is
small, we can neglect $\mathcal{O}(\lambda^{2})$ terms and find the
comp/lete state to be 
\[
|\psi^{(1)}\rangle:=U_{int}|\psi^{(0)}\rangle=|\psi_{S}\rangle|0,0\rangle-i\lambda\left(c_{i}|\psi_{S}\rangle|1,0\rangle+c_{j}|\psi_{S}\rangle|0,1\rangle\right)+\mathcal{O}(\lambda^{2})
\]
\item Next, the fermions in the left and right wire interfere in a beam
splitter. Written in the basis $\{|00\rangle,|01\rangle,|10\rangle,|11\rangle\}$
the beam splitter can in general be described by the scattering matrix
\[
S=\left(\begin{array}{cccc}
1\\
 & r' & t\\
 & t' & r\\
 &  &  & rr'-tt'
\end{array}\right),
\]
where $r$ and $t$ ($r'$ and $t'$) are the reflection and transmission
amplitudes for a fermion incident from the left (right) 
side. Though not important in our case, since the state $|1,1\rangle$ where
there is a fermion in each wire is suppressed by $\lambda^{2}$, lets
give a short explanation for how to to obtain last entry $|1,1\rangle\to(rr'-tt')|1,1\rangle$.
One has take into account that the wave function is antisymmetric,
$|1,1\rangle=|\phi_{L}\rangle\otimes|\phi_{R}\rangle-|\phi_{R}\rangle\otimes|\phi_{L}\rangle$.
Here the position in the tensor product labels the fermion (say they
are called $1$ and $2$), whereas $|\phi_{L}\rangle$ and $|\phi_{R}\rangle$
are single fermion states in the left and right wire. Then 
\begin{align*}
|1,1\rangle  \to  \left(r|\phi_{L}\rangle+t|\phi_{R}\rangle\right)\otimes\left(r'|\phi_{R}\rangle+t'|\phi_{L}\rangle\right)
 -\left(r'|\phi_{R}\rangle+t'|\phi_{L}\rangle\right)\otimes\left(r|\phi_{L}\rangle+t|\phi_{R}\rangle\right)
\end{align*}
 leads to the entry $(rr'-tt')$. Unitarity demands $|r|^{2}+|t|^{2}=1$,  $|r'|^{2}+|t'|^{2}=1$ and
$\bar{r}t'+\bar{t}r'=0$ (the condition $|rr'-tt'|^{2}=1$ is then
automatically fulfilled). 
\begin{itemize}
\item (a): Choosing a symmetric beam splitter, $r=r'$ and $t=t'$, allows
to measure the imaginary part of $G_{ij}$. Note that the unitary
constraints can now be expressed as $r=\sin\theta e^{i\varphi}$ and
$t=i\cos\theta e^{i\varphi}$ and we set the overall phase $\varphi=0$
since this will not change the result. The state evolves to
\begin{align*}
|\psi^{(2,a)}\rangle= &S^{(a)}|\psi^{(1)}\rangle\\
= &|\psi_{S}\rangle|0,0\rangle-i\lambda c_{i}|\psi_{S}\rangle\left(\sin\theta|1,0\rangle+i\cos\theta|0,1\rangle\right)
-i\lambda c_{j}|\psi_{S}\rangle\left(\sin\theta|0,1\rangle+i\cos\theta|1,0\rangle\right).
\end{align*}
\item (b): If the beam splitter is symmetric, except that transmitted fermions
incident from the right will accumulate an additional phase $\pi$,
i.e. $r=r'$ and $t=-t'$, this allows to measure the real part of $G_{ij}$. Note that the unitary constraints result it $r=\sin\theta e^{i\varphi}$
and $t=\cos\theta e^{i\varphi}$ and again we set $\varphi=0$. The
state evolves to
\begin{align*}
|\psi^{(2,b)}\rangle=&S^{(b)}|\psi^{(1)}\rangle\\
=&|\psi_{S}\rangle|0,0\rangle-i\lambda c_{i}|\psi_{S}\rangle\left(\sin\theta|1,0\rangle+\cos\theta|0,1\rangle\right)
-i\lambda c_{j}|\psi_{S}\rangle\left(\sin\theta|0,1\rangle-\cos\theta|1,0\rangle\right).
\end{align*}
\end{itemize}
\item Finally, we measure the particle number $n_{L}=c_{L}^{\dagger}c_{L}$
and $n_{R}=c_{R}^{\dagger}c_{R}$ in the left and right wire. Denoting
averages w.r.t the system $|\psi_{S}\rangle$ by $\langle...\rangle_{S}$
as in $G_{ij}=\langle c_{j}^{\dagger}c_{i}\rangle_{S}$ we find for
case (a)
\begin{align*}
\langle n_{L}\rangle^{(a)} & =\lambda^{2}\left(\sin^{2}\theta\langle n_{i}\rangle_{S}+\cos^{2}\theta\langle n_{j}\rangle_{S}-2\sin\theta\cos\theta\mathrm{\Im}(G_{ij})\right)\\
\langle n_{R}\rangle^{(a)} & =\lambda^{2}\left(\cos^{2}\theta\langle n_{i}\rangle_{S}+\sin\theta\langle n_{j}\rangle_{S}+2\sin\theta\cos\theta\Im(G_{ij})\right).
\end{align*}
Choosing an angle $\theta=\pi/4$ gives the imaginary part of $G_{ij}$,
\[
2\lambda^{2}\Im(G_{ij})=\langle n_{R}\rangle^{(a)}-\langle n_{L}\rangle^{(a)}.
\]
For the case (b), one gets
\begin{align*}
\langle n_{L}\rangle^{(b)} & =\lambda^{2}\left(\sin^{2}\theta\langle n_{i}\rangle_{S}+\cos^{2}\theta\langle n_{j}\rangle_{S}-2\sin\theta\cos\theta\mathrm{\Re}(G_{ij})\right)\\
\langle n_{R}\rangle^{(b)} & =\lambda^{2}\left(\cos^{2}\theta\langle n_{i}\rangle_{S}+\sin\theta\langle n_{j}\rangle_{S}+2\sin\theta\cos\theta\Re(G_{ij})\right).
\end{align*}
Choosing the same angle $\theta=\pi/4$ gives the real part of $G_{ij}$,
\[
2\lambda^{2}\Re(G_{ij})=\langle n_{R}\rangle^{(b)}-\langle n_{L}\rangle^{(b)}.
\]
\end{itemize}
Hence, both the real and imaginary parts of the coherence, as well as their statistical distribution, are experimentally measurable (at least in principle).

\section{Singular behaviour of non-connected correlation functions\label{app:Singular-behavior-of}}

For the example of $n=2$, we will show here why (\ref{eq:g_n^s_evolution})
produces solutions $g_{t}^{s}$ that are singular at coinciding points,
whereas solutions $g_{t}$ of (\ref{eq:g_n_evolution}) are continuous
(but not differentiable). The corresponding equations for $n=2$ are
(\ref{eq:g_2^s_evolution}) and (\ref{eq:g_2_evolution}). Namely,
\[
(\partial_{t}-\Delta)g_{t}^{s}(x,y)=2\partial_{x}\partial_{y}(\delta(x-y)g_{t}(x)g_{t}(y)),
\]
and
\[
(\partial_{t}-\Delta)g_{t}(x,y) =2\delta(x-y)\partial_{x}g_{t}(x)\partial_{y}g_{t}(y)).
\]

Integrating these equations across the diagonal line $\{x=y\}$ reveals
that the van Neumann boundary conditions will be singular for $g^{s}_t$
and regular for $g_t$. 

Let us do this explicitly for (\ref{eq:g_2^s_evolution}). We rotate
the variables $(x,y)$ by $\pi/4$ clockwise, $(v,u)=(\frac{x-y}{\sqrt{2}},\frac{x+y}{\sqrt{2}})$.
Then the derivative $\partial_{v}=\frac{\partial_{x}-\partial_{y}}{\sqrt{2}}$
is orthonormal to the line $\{x=y\}$ and $\partial_{u}=\frac{\partial_{x}+\partial_{y}}{\sqrt{2}}$
is parallel to the line. In these variables the equation reads
\[
(\partial_{t}-(\partial_{u}^{2}+\partial_{v}^{2}))g_{t}^{s}(\frac{u+v}{\sqrt{2}},\frac{u-v}{\sqrt{2}})=(\partial_{u}^{2}-\partial_{v}^{2})(\delta(\sqrt{2}v)\rho_{t}(\frac{u+v}{\sqrt{2}})\rho_{t}(\frac{u-v}{\sqrt{2}})).
\]
Integrating $\int_{-\epsilon}^{\epsilon}dv$ this equation and keeping
only terms of $\mathcal{O}(1)$,
\[
\partial_{v}g_{t}^{s}(\frac{u+v}{\sqrt{2}},\frac{u-v}{\sqrt{2}})\Big|_{-\epsilon}^{\epsilon}=\partial_{v}(\delta(\sqrt{2}v)\rho_{t}(\frac{u+v}{\sqrt{2}})\rho_{t}(\frac{u-v}{\sqrt{2}}))\Big|_{-\epsilon}^{\epsilon},
\]
The expression simplifies due to cyclic invariances, $g_{t}^{s}(x,y)=g_{t}^{s}(y,x)$,
\[
2\partial_{v}g_{t}^{s}(\frac{u+v}{\sqrt{2}},\frac{u-v}{\sqrt{2}})\Big|_{v=\epsilon}=2\sqrt{2}\delta'(\sqrt{2}v)\rho_{t}(\frac{u+v}{\sqrt{2}})\rho_{t}(\frac{u-v}{\sqrt{2}}))\Big|_{v=\epsilon}
\]
 Finally we take $\epsilon\to0^{+}$ and find the van Neumann boundary
condition in both sectors $\{x>y\}$ and $\{x<y\}$,
\[
\partial_{v}g_{t}^{s}|_{x=y^{+}}=-\partial_{v}g_{t}^{s}|_{x=y^{-}}=\sqrt{2}\delta'(0^{+})\rho(x)^{2}.
\]
The delta function will make this derivative blow up and therefore
$g_{t}^{s}(x,y)$ is indeed singular at $x=y$. The same procedure
for (\ref{eq:g_2_evolution}) yields
\begin{equation}\label{eq:g_2_neuman_conditions}
\partial_{v}g_{t}|_{x=y^{+}}=-\partial_{v}g_{t}|_{x=y^{-}}=-\frac{1}{\sqrt{2}}\rho_{t}'(x)^{2},
\end{equation}
which is finite.

\section{Solution for the discrete density\label{app:Derivation-density}}

The time evolution equation for the discrete density (\ref{eq:density_discrete})
can be rewritten as $\partial_{t}n=An+n$, with $n=(n_{1},...,n_{L})^{T}$,
$b=(\alpha_{1},...,\alpha_{L})^{T}$ and 
\[
A=\left(\begin{array}{ccccc}
-1-(\alpha_{1}+\beta_{1}) & 1\\
1 & -2 & ...\\
 & ... & ... & ...\\
 &  & ... & -2 & 1\\
 &  &  & 1 & -1-(\alpha_{L}+\beta_{L})
\end{array}\right).
\]
If $A$ is diagonalisable, $A=SDS^{-1}$, a general solution for an
initial condition $n(0)=u$ is given by
\begin{align}
n(t) & =\exp(At)u+(\exp(At)-1)A^{-1}b\nonumber \\
 & =S\exp(Dt)(S^{-1}u+D^{-1}S^{-1}b)+A^{-1}b\label{eq:general_solution_vector_DE}
\end{align}

If $\alpha_{1}+\beta_{1}=1=\alpha_{L}+\beta_{L}$, $A$ has is a so-called
Tölpitz matrix and has eigenvalues $\lambda_{k}$ and normalised eigenvectors
$v_{k}$,
\begin{align*}
\lambda_{k} & =-2+2\cos(k)\\
(v_{k})_{j} & =\sqrt{2/(L+1)}\sin(jk)
\end{align*}
 with $k=\frac{\pi}{L+1},\frac{2\pi}{L+1},...,\frac{L\pi}{L+1}$,
from which (\ref{eq:density_discrete_solution}) can be derived. For general
injection/extraction rates we make an ansatz
\begin{align*}
\lambda_{k} & =-2+2\cos(k)\\
(v_{k})_{j} & =e^{ijk}+s_{k}e^{-ijk}
\end{align*}
where $s_{k}$ and $k$ are determined by the eigenvalue equations.
Denoting $z:=e^{ik}$ this leads to
\begin{align}
0 & =(z^{L+1}-z^{-(L+1)})-(A+B)(z^{L}-z^{-L})+AB(z^{L-1}-z^{-(L-1)})\label{eq:quantization_equation}\\
s_{k} & =-e^{2ik}\frac{A-e^{ik}}{A-e^{-ik}},\nonumber 
\end{align}
where $A=1-(\alpha_{1}+\beta_{1})$ and $B=1-(\alpha_{L}+\beta_{L})$.
Note that even though this equation has $2L+2$ solutions for $z$,
only $L$ on them give rise to eigenvalues with linearly independent
eigenvectors: If $k$ is a solution then also $-k$ is a solution,
but the corresponding eigenvectors are linearly dependent, $v_{-k}=s_{k}v_{k}$.
Furthermore $k=0$ is a solution with zero eigenvector. Numerical
solutions of \ref{eq:quantization_equation} show that for $A,B\in[1,-1]$
all solutions of $z$ lie on the unit circle and therefore (taking
into account that $k\sim-k$ and $k\ne0$) we have $k\in(0,\pi)$. 

A site in the bulk, $j\sim aL$, approaches the steady value $(A^{-1}b)_{j}$
according to the time dependent part of (\ref{eq:general_solution_vector_DE}).
Again, since $\lambda_{k}<0$, there is an exponential decay. Only
the smallest solution for $k$, denoted by $k^{*}$, will contribute. In the
limit of large system-size $L$ one can check that $k^{*}\approx\pi/L$
is the smallest positive solution for $k$. Note that the amplitude
$(v_{k^{*}})_{j}\approx e^{ia\pi}-e^{-ia\pi}$ of this term stays
finite ($s_{k^{*}}\approx-1$). Then, the time scale with which
the bulk approaches the steady value is, as before, $t_{\mathrm{decay}}\sim\mathcal{O}(L^{2})$.

At the boundary, the amplitude of the term corresponding to $k^{*}$
will be zero. E.g. for $j=1$, $(v_{k^{*}})_{1}=e^{ik^{*}}+s_{k^{*}}e^{-ik^{*}}=0$.
To have a non-zero amplitude, one has to consider terms where $k\sim bL$
with $b\sim\mathcal{O}(1)$, which leads to $t_{decay}\sim\mathcal{O}(1)$.

\section{Analytic solution of the connected 2-point function\label{app:Analytic-solution-of}}
Here we outline how to solve the connected 2-point function analytically,
which was needed for the comparison with the solution of the discrete
equations in the scaling limit in Fig. (\ref{fig:g_2 with fitting initial})
and (\ref{fig:g_2 with unfitting initial}).

\paragraph{Solution for the density.}

First we construct an analytic solution of the density (\ref{eq:density_scaling})
with domain wall initial condition $\rho(x,0)=\Theta(1/2-x)$ and
boundary conditions $\rho(0,t)=n_{a},\rho(1,t)=n_{b}.$ We simplify
the boundary conditions by subtracting the stationary solution, $\rho_{\infty}(x)=n_{a}+x(n_{b}-n_{a})$.
Then we solve for $\tilde{\rho}(x,t)=\rho(x,t)-\rho_{\infty}(x)$,
which has easier boundary conditions.
\begin{align*}
(\partial_{t}-\partial_{x}^{2})\tilde{\rho} & =0\\
\tilde{\rho}(x,0) & =\Theta(1/2-x)-n_{a}-x(n_{b}-n_{a})\\
\tilde{\rho}(0,t) & =\tilde{\rho}(1,t)=0
\end{align*}
We find the solution by an expansion in $\{\sin(n\pi x)\}_{n=1}^{\infty}$.
These functions satisfy the correct boundary conditions and are orthogonal
in the sense that $\int_{0}^{1}\sin(n\pi x)\sin(m\pi x)dx=\frac{1}{2}\delta_{nm}$.
Importantly, they form a complete basis of $L^{2}([0,1])$ which justifies
the expansion. 

Taking into account the initial condition, this leads to 
\begin{align*}
\tilde{\rho}(x,t) & =\sum_{n=1}^{\infty}c_{n}\sin(n\pi x)e^{-n^{2}\pi^{2}t}\\
c_{n} & =\frac{2}{n\pi}(1-(-1)^{n}(n_{a}-n_{b})-2n_{a}\delta_{n,odd}-(-1)^{n/2}\delta_{n,even}).
\end{align*}
In the special case where the boundary conditions match the initial
conditions ($n_{a}=1$, $n_{b}=0$) one finds

\begin{align*}
\rho(x,t) & =1-x-\sum_{k=1}^{\infty}\frac{(-1)^{k}}{k\pi}\sin(2\pi kx)e^{-4\pi^{2}k^{2}t}.
\end{align*}

\paragraph{Solution for the connected two point function.}

In appendix \ref{app:Singular-behavior-of} we saw that the connected
two point function $g_{t}(x,y)$ satisfies the van-Neumann boundary
conditions (\ref{eq:g_2_neuman_conditions}) on the diagonal $\{x=y\}$.
Here we construct a solution of (\ref{eq:g_2_evolution}) on the lower
triangle $T^{+}=\{(x,y)\in[0,1]^{2}\colon x\geq y\}$: First we identify
a function that satisfies the boundary conditions,
\[
w(x,y,t)=y(1-x)\partial_{x}\rho(x,t)\partial_{y}\rho(y,t).
\]
Note that for $t\to\infty$, $w$ becomes the correct stationary solution
on $T^{+}$. Then we solve for $f(x,y,t):=g_{t}(x,y)-w(x,y,t)$, which
satisfies a inhomogeneous heat equation with homogeneous boundary
conditions
\begin{align*}
(\partial_{t}-\Delta)f(x,y,t) & =S(x,y,t):=2(1-x)\partial_{x}\rho(x,t)\partial_{y}^{2}\rho(y,t)-2y\partial_{x}^{2}\rho(x,t)\partial_{y}\rho(y,t)\\
f(x,y,0) & =-y(1-x)\delta(1/2-x)\delta(1/2-y)\\
f(x,0,t) & =0=g(1,0,t)\text{ (Dirichlet condition)}\\
\partial_{v}f|_{x=y} & =0\text{ (Neumann condition)}
\end{align*}
with $\partial_v :=(\partial_x-\partial_y)/\sqrt{2}$ as in appendix \ref{app:Singular-behavior-of}. This is solved, as before, by the method of eigenfunction expansion.
Note that 
\[
\psi_{nm}(x,y):=\sin(n\pi x)\sin(m\pi y)+\sin(m\pi x)\sin(n\pi y)
\]
 is a complete basis of $L^{2}(T^{+})$ that satisfies the correct
Dirichlet and Neumann boundary conditions. It satisfies 
\[
\int_{T^{+}}\psi_{nm}(x,y)\psi_{kl}(x,y)dxdy=\frac{\delta_{nk}\delta_{ml}+\delta_{nl}\delta_{mk}}{4}.
\]
We write $S(x,y,t)=\sum_{n\ge m\ge1}\hat{S}_{nm}(t)\psi_{nm}(x,y)$
and $f(x,y,t)=\sum_{n\ge m\ge1}\hat{f}_{nm}(t)\psi_{nm}(x,y)$, where
the coefficients are given by
\[
\hat{S}_{nm}(t)=\begin{cases}
4\int_{T^{+}}S\psi_{nm} & \text{if }n>m\\
2\int_{T^{+}}S\psi_{nn} & \text{if }n=m
\end{cases}.
\]
This leads to the $\partial_{t}\hat{f}_{nm}+\pi^{2}(n^{2}+m^{2})\hat{f}_{nm}=\hat{S}_{nm}$
which is solved by 
\[
\hat{f}_{nm}(t)=\underbrace{\hat{f}_{nm}(0)e^{-\pi^{2}(n^{2}+m^{2})t}}_{=:\hat{f}_{nm}^{hom}(t)}+\underbrace{\int_{0}^{t}e^{\pi^{2}(n^{2}+m^{2})(\tau-t)}\hat{S}_{nm}(\tau)}_{=:\hat{f}_{nm}^{part}(t)}.
\]
We get a solution $f_{hom}=\sum_{n\ge m\ge1}\hat{f}_{nm}^{hom}(t)\psi_{nm}$
of the homogeneous equation that satisfies the initial condition and
a solution $f_{part}=\sum_{n\ge m\ge1}\hat{f}_{nm}^{part}(t)\psi_{nm}$.
One finds
\begin{align*}
f_{hom}(x,y,t) & =\sum_{k,l\ge0}(-1)^{k+l+1}e^{-\pi^{2}((2k+1)^{2}+(2l+1)^{2})t}\sin((2k+1)\pi x)\sin((2l+1)\pi y)\\
 & =-\frac{1}{4}\vartheta_{1}(\pi x,e^{-4\pi^{2}t})\vartheta_{1}(\pi y,e^{-4\pi^{2}t}).
\end{align*}
The precise expression for $\hat{S}_{nm}$ is rather complicated,
and it is easier to solve for $f_{part}$ using the Mathematica NDSolve
function. The complete solution is then $g_{t}(x,y)=w(x,y,t)+f_{hom}(x,y,t)+f_{part}(x,y,t)$.
By symmetry in $x$ and $y$, i.e. $g_{t}(x,y)=g_{t}(y,x)$, this
also determines a solution on $T^{-}:=\{(x,y)\in[0,1]^{2}\colon x\geq y\}$.

\section{Proofs}

\subsection{Expansion of loop expectation values into non-crossing partitions}\label{app:non-crossing_partitions}
The proof of \eqref{eq:moment-cumulant_Q-SSEP} and its continuous version (\ref{eq:g^s_expansion_in_g}), which relate the correlation function $\mathbb E[G_{i_1,i_2}\cdots G_{i_n,i_1}]=:[n]$ to its connected part $\mathbb E[G_{i_1,i_2}\cdots G_{i_n,i_1}]^c=:[n]^c$ through a sum over non-crossing partitions (the notation $[n]$ and $[n]^c$ is only used in this subsection) is based on the three conditions \eqref{eq:u(1)-invariance}-\eqref{eq:factorisation_of_loops} which can be stated more compactly as
\begin{itemize}
    \item The measure $\mathbb E$ is $U(1)$ invariant (we will often denote $\mathbb E[\cdots]=[\cdots]$ where we dropped the subscript $t$ from $\mathbb E_t$).
    \item Connected loop expectation value of $n$ points scale as $[n]^c\sim\frac {1}{L^{n-1}}$ and all other connected correlations with equal number are either of the same order or sub-leading.
\end{itemize}
Using the moment-cumulant formula (\ref{eq:moments_as_cumulants}), the essential point is to show that cumulants 
\begin{equation}
\mathbb E_\pi[G_{i_1 i_2}\cdots G_{i_n i_1}]^c:=\prod_{b\in \pi}\mathbb E[G_{i_{b(1)}i_{b(2)}}...G_{i_{b(|b|)}i_{b(1)}}]^c
\end{equation}
corresponding to crossing partitions $\pi$ will be sub-leading compared to non-crossing partitions $\pi$ in the scaling limit.
\paragraph{Scaling of non-crossing partitions.}
We start the proof by showing that if $\pi=\{b^{(1)},\cdots,b^{(m)}\}$ is a non-crossing partition of $\{12,23\cdots,n1\}$, i.e.\ of the edges of a loop with $n$ nodes, then $[\pi]^c:=\mathbb E_\pi[G_{i_1 i_2}\cdots G_{i_n i_1}]^c$ scales as $L^{-n+1}$ independently of the number of blocks $m$ -- and therefore behaves in the same way as the single loop expectation value $\mathbb E[G_{i_1 i_2}\cdots G_{i_n i_1}]^c$ with $n$ points.

First notice, that if the blocks of $\pi$ are not nested into each other (e.g $\pi=\{\{12,23\},\{34\},\{45,56,61\}\}$) then, by $U(1)$ invariance, we have to connect starting and endpoints of each block by a Kronecker $\delta$. Once we arrive at the last block, this conditions is already satisfied due to the other delta functions. We therefore need $m-1$ delta functions,
\begin{align*}
[\pi]^c=\underbrace{\delta \cdots \delta}_{m-1} \ [b^{(1)}]^c\cdots  [b^{(m)}]^c
&\sim L^{-m+1} L^{-\sum_{i=1}^m |b^{(i)}|+m}\\
&\sim L^{-n+1},
\end{align*}
where $\delta\sim L^{-1}$ in the scaling limit and the sum over the size $|b^{(i)}|$ of each block is equal to the total number of elements $n$.

Next, assume that $\pi$ has some nested blocks (e.g. $\pi=\{\{12,23\},\{45\},\{34,56,61\}\}$), then treat each collection of nested blocks as a big block $B$ (e.g. $B=\{\{45\},\{34,56,61\}\}$), such that the argument above applies to the non-nested blocks and the big blocks. Now we can iterate the argument for each big block $B$ and possible collections of nested sub-blocks therein. In the end we are left with $m-1$ delta functions needed to close each block in $\pi$ to form a loop. This shows that $[\pi]^c\sim L^{-n+1}$ if $\pi$ is non-crossing.

\paragraph{Scaling of crossing partitions.}
Now assume that $\pi$ consists of a collection $B$ of non-crossing blocks with $|B|$ elements in total and a collection $C$ of crossing blocks that cannot be disentangled from another with $|C|$ elements in total (e.g. $\pi=\{\{12,34\},\{23,45\},\{56\},\{67\}\}$, $B=\{\{56\},\{67\}\}$ and $C=\{\{12,34\},\{23,45\}\}$). Then we have $|B|+|C|=n$. If we treat $B$ as an independent partition of the cyclic set formed by all its elements, then by the argument above, $[B]^c\sim L^{-|B|+1}$. Alternatively we can view $B$ as the partition of the full loop from which we removed all edges that belong to the collection $C$ of crossing blocks, leaving us with a smaller loop of $|B|$ edges. If we assume that $[C]^c\sim L^{-|C|}$ (as will be shown below) then
\begin{align*}
    [\pi]^c=\delta \ [B]^c[C]^c
    &\sim L^{-1}L^{-|B|+1}L^{-|C|}\\
    &\sim L^{-n},
\end{align*}
where only a single delta function is needed to connect the collection $B$ and $C$ (because all the other delta functions are already included in $[B]^c$ and $[C]^c$. If $\pi$ has more than one collection $C$ of crossing blocks that cannot be disentangled, then it will scale with even higher negative power of $L$. This shows that crossing partitions are sub-leading in the scaling limit compared to non-crossing partitions.

It remains to show that $[C]^c\sim L^{-|C|}$ (probably it is even true that $[C]^c \sim L^{-|C|-1}$). The idea is to produce a collection of crossing blocks starting with non-crossing blocks and permuting its elements. Assume that $\{b^{(1)},b^{(2)}\}$ is a collection of two non-crossing blocks 
\begin{align*}
    b^{(1)}&=\{i_1 i_2,\cdots,i_{k} i_{k+1}\} & b^{(2)}&=\{i_{k+1} i_{k+2},\cdots,i_n i_1\}
\end{align*}
which would mean that there is only one delta function $\delta(i_1,i_{k+1})$ necessary for the product of these blocks to be non-zero. Then construct a first crossing by inserting an element $i_l i_{l+1}$ from the "middle" of $b^{(1)}$ (and not from the boundary) into $b^{(2)}$,
\begin{align*}
    b^{(1)}&\to b^{(1)}=\{i_1 i_2,\cdots,i_{l-1} i_l,i_{l+1} i_{l+2},\cdots,i_{k} i_{k+1}\}& 
    b^{(2)}&\to b^{(2)}=\{i_l i_{l+1},i_{k+1} i_{k+2},\cdots,i_n i_1\},
\end{align*}
which makes it necessary to insert a second delta function, $\delta(i_l,i_{l+1})$. In this way one can continue to permute elements between the two blocks, create more crossings and obtain $C=\{b^{(1)},b^{(2)}\}$ of sizes $|b^{(1)}|=k'$ and $|b^{(2)}|=n-k'$. Each new crossing necessitates a new delta function. Note, however, that a new crossing is only created, if one permutes an element that (a) is still in its original block, (b) whose neighbours have not yet been permuted (we sort the elements $i_l i_{l+1}$ in ascending order with respect to the index $l$) and (c) is not taken from the boundary of the block. Since no connected correlation function is more dominant than that of single loops, the two blocks, after an arbitrary permutation between them, will scale at most as  $[b^{(1)}]\sim L^{-k'+1}$ and $[b^{(2)}]\sim L^{-(n-k')+1}$. We therefore have
\begin{align*}
    [C]^c&=\underbrace{\delta \cdots \delta}_{\#\text{crossings}+1} [b^{(1)}]^c \, [b^{(2)}]^c\\
    &\sim L^{-n+1-\#\text{crossings}}.
\end{align*}
As a consequence, in the case where $C$ consists of two crossing blocks, this shows that it scales at most with $[C]^c\sim L^{-|C|}$ where $|C|=n$ in our example. 

If $C$ is a collection of $m$ crossing blocks it is probably still true that the number of delta functions is equal to $\#\text{crossings}+1$. However, all we need here is that $m$ crossing blocks will cause at least $m$ delta functions. Then the scaling is
\begin{align*}
    [C]^c&=\underbrace{\delta \cdots \delta}_{m} \ [b^{(1)}]^c\cdot[b^{(m)}]^c\\
    &\sim L^{-|C|}.
\end{align*}
is sufficient. We saw that $2$ crossing blocks cause at least $2$ delta functions. But each originally non-crossing block that we add to the collection of crossing blocks already comes with one delta function, even before we permute its elements with the other blocks. For example, if we have the crossing blocks $\{\{12,34\},\{23,41\}\}$ which need $2$ delta functions $\delta(2,3)\delta(1,4)$ and we add a block, we have $\{\{12,34\},\{23,45\},\{56,61\}$, which needs $3$ delta functions, $\delta(2,3)\delta(1,4)\delta(4,5)$. If we now cross the new block with the others this cannot reduce the number of delta functions. Therefore, this argument shows that $m$ crossing blocks indeed cause $m$ delta functions and concludes the proof.

\subsection{Time evolution of connected loop expectation values}\label{app:derivation-connected-correlations}

We will show how (\ref{eq:g_n_evolution}) follows from (\ref{eq:g_n^s_evolution})
by induction over $n$. We start by writing the sum $2\sum_{i<j}=\sum_{ij,j\neq i}$
in these equations as a sum over all ordered, non-crossing and non-empty
sets $r=\{r_{1},r_{2},...\}$ and $s=\{s_{1},s_{2},...\}$ with $i=r_{1}$
and $j=s_{1}$ such that $r\sqcup s=[n]\equiv\{x_{1},...,x_{n}\}$. Throughout this paper we use the symbol $\sqcup$ to denote the union of ordered, non-crossing and non-empty subsets. 
Furthermore, instead of $g_{t}(x_{i},x_{i+1},...,x_{j-1})$ we simply
write $g(r)$ if $r=\{x_{i},x_{i+1},...,x_{j-1}\}$. Then the two equations
become,
\begin{equation}
(\partial_{t}-\Delta)g_{t}^{s}([n])=\sum_{r\sqcup s=[n]}\partial_{r_{1}}\partial_{s_{1}}(\delta(r_{1},s_{1})g_{t}^{s}(r)g_{t}^{s}(s)),\label{eq:g^s_evolution-app}
\end{equation}
and
\begin{equation}
(\partial_{t}-\Delta)g_{t}([n])=\sum_{r\sqcup s=[n]}\delta(r_{1},s_{1})\partial_{r_{1}}g_{t}(r)\partial_{s_{1}}g_{t}(s).\label{eq:g_evolution-app}
\end{equation}

For $n=1$ the two equations are identical. Let's assume that (\ref{eq:g_evolution-app})
holds for all $k\le n-1$ for some $n\in\mathbb{N}$. We will use
(\ref{eq:g^s_evolution-app}) to show that it holds also for $k=n$. 

From (\ref{eq:g^s_expansion_in_g}) we know that the $g_{t}^{s}$ can
be expanded into $g_{t}$ as a sum over non-crossing partitions $\pi$ of $[n]=\{x_1,\cdots,x_n\}$. Also recall that the dual partition $\pi^*$ is a partition on the edges, which we name according to the convention
\begin{equation}\label{eq:convention}
D_{\pi}=
	\raisebox{-0.5\height}{\includegraphics{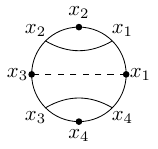}}.
\end{equation}
In this example $\pi=\{\{x_{1},x_{2}\},\{x_{3},x_{4}\}\}$ and $\pi^{*}=\{\{x_{1},x_{3}\},\{x_{2}\},\{x_{3}\}\}$.
Then, denoting $\delta(d)=\delta(d_1,\cdots,d_{|d|})$ where $d$ is a block in $\pi^*$, the expansion in non-crossing partitions is
\begin{equation}
g^{s}_t([n])=\sum_{\pi\in NC([n])}\prod_{d\in\pi^{*}}\delta(d)\prod_{b\in\pi}g_t(b).\label{eq:g^s_expansion_in_g-app}
\end{equation}
With $D_\pi:=\prod_{d\in\pi^{*}}\delta(d)\prod_{b\in\pi}g_t(b)$ we can evaluate the right hand side of (\ref{eq:g^s_evolution-app})
\begin{align}
rhs & =\sum_{r\sqcup s=[n]}\partial_{r_{1}}\partial_{s_{1}}\Big(\delta(r_{1},s_{1})\sum_{\rho\in NC(r)}D_{\rho}\sum_{\sigma\in NC(s)}D_{\sigma}\Big)\label{eq:rhs}\\
 & =\sum_{\pi\in NC([n])\backslash[n]}\sum_{\underset{(x,y)\in d}{d\in\pi^{*}}}\partial_{x}\partial_{y}D_{\pi},\nonumber 
\end{align}
where $(x,y)\in d$ denotes all tuples with $x\neq y$. To see the
second equal sign, one has to establish the bijection
\begin{gather*}
\{\delta(r_{1},s_{1})D_{\rho}D_{\sigma}\mid\rho\in NC(r),\sigma\in NC(s),s\sqcup r=[n]\}\\
\longleftrightarrow\\
\{(D_{\pi},x,y)\mid\pi\in NC([n])\backslash[n],(x,y)\in d\text{ st. }d\in\pi^{*}\}.
\end{gather*}

\begin{enumerate}
\item [{``$\rightarrow$'':}] One draws the two diagrams $D_{\rho}$
and $D_{\sigma}$ inside a single big loop with nodes $\{r_{1},r_{2},...,s_{1},s_{2},...\}=[n]$
and connects $r_{1}$ and $s_{1}$ by a dotted line to represent the
delta function $\delta(r_{1},s_{1})$. This is equal to $D_{\pi}$
with two marked nodes $x=r_{1}$ and $y=s_{1}$, where $\pi=\rho\cup\sigma$
is a partition on the big circle with at least two elements -- therefore
excluding the partition $\pi=\{[n]\}$.
\item [{``$\leftarrow$'':}] One starts with $(D_{\pi},x,y)$. Since
the two marked nodes $(x,y)\in d\in\pi^{*}$ belong to a block of
the dual partition, one can cut the diagram at these two nodes without
breaking any block $b\in\pi$. The cut therefore defines independent
non-crossing partitions $\rho$ and $\sigma$ on the two non-crossing
subsets $r\sqcup s=[n]$ where $r_{1}=x$ and $s_{1}=y$.
\begin{equation}\label{eq:bijection}
D_{\pi}=
	\raisebox{-0.5\height}{\includegraphics{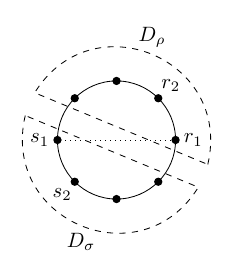}}
\end{equation}
\end{enumerate}
With the help of (\ref{eq:rhs}) and (\ref{eq:g^s_expansion_in_g-app}) we
can rewrite (\ref{eq:g^s_evolution-app}) and isolate the connected
correlation function $g([n])$, which is the term we are aiming for,
\begin{equation}
(\partial_{t}-\Delta)g_t([n])=\sum_{\pi\in NC([n])\backslash[n]}\left(-(\partial_{t}-\Delta)D_{\pi}+\sum_{\underset{(x,y)\in d}{d\in\pi^{*}}}\partial_{x}\partial_{y}D_{\pi}\right).\label{eq:g_calculation}
\end{equation}
The only information which is missing is the action of $(\partial_{t}-\Delta)$
on $D_{\pi}$. We claim that
\begin{equation}
(\partial_{t}-\Delta)D_{\pi}=\sum_{\underset{d\in\pi^{*}}{(x,y)\in d}}(\partial_{x}\partial_{y}-\partial_{x}^{g}\partial_{y}^{g})D_{\pi}+\sum_{\underset{b\in\pi}{(x,y)\in b}}\partial_{x}^{g}\partial_{y}^{g}D_{\pi\backslash b\cup b_{1}(x,y)\cup b_{2}(x,y)},\label{eq:action_on_D}
\end{equation}
the derivation of which comes at the end of this section. Let us explain
the arising terms:
\begin{itemize}
\item The first term is a sum over all blocks $d$ of the dual partition
$\pi^{*}$, from which we choose all possible tuples $(x,y)$ with $x\neq y$. The symbol $\partial_{x}^{g}$ means that the derivative
only acts on the $g$'s and not the delta functions that appear in
$D_{\pi}$, while $\partial_{x}$ acts on both.
\item The second term is a sum over all blocks $b\in\pi$, from which we
choose all possible tuples of edges $(x,y)$ with $x\neq y$. By the convention (\ref{eq:convention}),
we denote the neighbouring nodes by the same name. This allows us to
cut the block $b$ of edges along the nodes $(x,y)$. The two resulting
blocks are denoted by $b_{1}(x,y)$ and $b_{2}(x,y)$. The partition
$\pi\backslash b\cup b_{1}(x,y)\cup b_{2}(x,y)$ is the one where
$b$ was removed and replaced by the two blocks $b_{1}(x,y)$ and $b_{2}(x,y)$.
\end{itemize}
Note that the term with $\partial_{x}\partial_{y}$ is cancelled once
we plug (\ref{eq:action_on_D}) into (\ref{eq:g_calculation}). This
equation simplifies further if we do the sum over all $\pi\in NC([n])\backslash[n]$,
\begin{equation}
\sum_{\pi\in NC([n])\backslash[n]}\left(-\sum_{\underset{d\in\pi^{*}}{(x,y)\in d}}\partial_{x}^{g}\partial_{y}^{g}D_{\pi}+\sum_{\underset{b\in\pi}{(x,y)\in b}}\partial_{x}^{g}\partial_{y}^{g}D_{\pi\backslash b\cup b_{1}(x,y)\cup b_{2}(x,y)}\right)=-\sum_{r\sqcup s=[n]}\partial_{r_{1}}^{g}\partial_{s_{1}}^{g}D_{\{r,s\}}.\label{eq:simplification}
\end{equation}
Let us explain this: Since $\pi\neq\{[n]\}$ one can always find a
block $d\in\pi^{*}$ that consists of at least two nodes $(x,y)$.
We can join the two blocks $b(x)\in\pi$ and $b(y)\in\pi$ to which
the corresponding edges $x$ and $y$ belong. This forms a block $b$
of a new partition $\pi'$ which differs from $\pi$ only by this
block, $\pi'\backslash b\cup b(x)\cup b(y)=\pi$. Note, that if $\pi$
consists of only two blocks, then $\pi'$ will consist of a single
block, so $\pi'=\{[n]\}$. There is hence a bijection 
\begin{gather*}
\{(D_{\pi},x,y)\mid\pi\in NC([n])\backslash[n],(x,y)\in d\text{ st. }d\in\pi^{*}\}\\
\leftrightarrow\\
\{\pi'\backslash b\cup b_{1}(x,y)\cup b_{2}(x,y)\mid\pi'\in NC[n],(x,y)\in b\in\pi'\}.
\end{gather*}
But $\pi'=[n]$ is not available in the above sum, therefore leaving
all those terms $$-\sum_{\underset{d\in\pi^{*}}{(x,y)\in d}}\partial_{x}^{g}\partial_{y}^{g}D_{\pi}$$
uncancelled, where $\pi$ consists of only two blocks. Then we can
write $\pi=\{s,r\}$ where $r\sqcup s=[n]$, which
establishes the above equality.

Equation (\ref{eq:g_calculation}) then simplifies to
\begin{equation}
(\partial_{t}-\Delta)g_t([n])=\sum_{r\sqcup s=[n]}\partial_{r_{1}}^{g}\partial_{s_{1}}^{g}D_{\{r,s\}}=\sum_{r\sqcup s=[n]}\delta(r_{1},s_{1})\partial_{r_{1}}g_t(r)\partial_{s_{1}}g_t(s),\label{eq:result_g_evolution}
\end{equation}
which is what we wanted to show. 

\paragraph*{Proof of (\ref{eq:action_on_D}).}

Recall that our notation $(x,y)\in b$, for $b$ any block of a partition, assumes that $x\not= y$.
Two identities about $\delta$-functions of several variables that
we will need are 
\begin{align}
\sum_{x\in[n]}\partial_{x}\delta([n])= & 0\label{eq:delta_identity_1}\\
\Delta_{[n]}\delta([n])= & -\sum_{\underset{x\neq y}{(x,y)\in[n]}}\partial_{x}\partial_{y}\delta([n])\label{eq:delta_identity_2}
\end{align}
The subscript of the laplacian denotes the set on which it acts. For
example $\Delta_{b}=\sum_{x\in b}\partial_{x}^{2}$ for some set $b\subset[n]$.
For $\pi\in NC([n])\backslash[n]$ one finds,
\begin{align*}
(\partial_{t}-\Delta)D_{\pi}= & \underbrace{-\left(\Delta\prod_{d\in\pi^{*}}\delta(d)\right)\prod_{b\in\pi}g_{t}(b)-2\sum_{x\in[n]}\left(\partial_{x}\prod_{d\in\pi^{*}}\delta(d)\right)\left(\partial_{x}\prod_{b\in\pi}g_{t}(b)\right)}_{(I)}\\
 & +\underbrace{\prod_{d\in\pi^{*}}\delta(d)(\partial_{t}-\Delta)\prod_{b\in\pi}g_{t}(b)}_{(II)}.
\end{align*}
Using $\sum_{x\in[n]}=\sum_{e\in\pi^{*}}\sum_{x\in e}$,
\[
(I)=-\sum_{e\in\pi^{*}}\prod_{d\in\pi^{*}\backslash e}\delta(d)\left(\left(\Delta_{e}\delta(e)\right)\prod_{b\in\pi}g_{t}(b)+2\sum_{x\in e}\partial_{x}\delta(e)\partial_{x}\prod_{b\in\pi}g_{t}(b)\right).
\]
With the help of the $\delta$-function identities (\ref{eq:delta_identity_1})
and (\ref{eq:delta_identity_2}),
\[
(I)=\sum_{e\in\pi^{*}}\prod_{d\in\pi^{*}\backslash e}\delta(d)\left(\sum_{(x,y)\in e}\left(\partial_{x}\partial_{y}\delta(e)\right)\prod_{b\in\pi}g_{t}(b)+2\sum_{(x,y)\in e}\partial_{y}\delta(e)\partial_{x}\prod_{b\in\pi}g_{t}(b)\right).
\]
This expression can be written as a total derivative of $\partial_{x}\partial_{y}$
up to the missing term $\partial_{x}^{g}\partial_{y}^{g}$,
\[
(I)=\sum_{e\in\pi^{*}}\prod_{d\in\pi^{*}\backslash e}\delta(d)\sum_{(x,y)\in e}\left(\partial_{x}\partial_{y}\left(\delta(e)\prod_{b\in\pi}g_{t}(b)\right)-\delta(e)\partial_{x}\partial_{y}\prod_{b\in\pi}g_{t}(b)\right),
\]
which, rewritten in terms of $D_{\pi}$, is the first term that appears
in (\ref{eq:action_on_D}),
\[
(I)=\sum_{e\in\pi^{*}}\sum_{(x,y)\in e}(\partial_{x}\partial_{y}-\partial_{x}^{g}\partial_{y}^{g})D_{\pi}.
\]

For the second term we need to use (\ref{eq:g_evolution-app}), which
by assumption holds for all $k\le n-1$. Since $\pi\in NC([n])\backslash[n]$ the assumption applies,
\begin{align*}
(II)= & \prod_{d\in\pi^{*}}\delta(d)\sum_{c\in\pi}(\partial_{t}-\Delta_{c})g_{t}(c)\prod_{b\in\pi\backslash c}g_{t}(b)\\
= & \prod_{d\in\pi^{*}}\delta(d)\sum_{c\in\pi}\sum_{r\sqcup s=c}\delta(r_{1},s_{1})\partial_{r_{1}}g_{t}(r)\partial_{s_{1}}g_{t}(s)\prod_{b\in\pi\backslash c}g_{t}(b).
\end{align*}
Instead of summing over $r\sqcup s=c$, we can also sum over $(x,y)\in c$
(where as usual $x\neq y$). In this case $r$ and $s$ are the blocks
$b_{1}(x,y)$ and $b_{2}(x,y)$ that result from cutting $c$ along
the nodes $(x,y)$,
\[
(II)=\prod_{d\in\pi^{*}}\delta(d)\sum_{c\in\pi}\sum_{(x,y)\in c}\delta(x,y)\partial_{x}g_{t}(b_{1}(x,y))\partial_{y}g_{t}(b_{2}(x,y))\prod_{b\in\pi\backslash c}g_{t}(b).
\]
The expression can therefore be rewritten in terms of the partition
$\pi\backslash c\cup b_{1}(x,y)\cup b_{2}(x,y)$ and provides the
second term in (\ref{eq:action_on_D}),
\[
(II)=\sum_{c\in\pi}\sum_{(x,y)\in c}\partial_{x}^{g}\partial_{y}^{g}D_{\pi\backslash c\cup b_{1}(x,y)\cup b_{2}(x,y)}.
\]
This concludes the derivation of the time evolution equation of connected
correlation functions.

\subsection{Time evolution of the new measure}\label{app:derivation-new-measure}

Here, we present the proof of the evolution equation (\ref{eq:phi_equation}) of the new measure $\varphi_t$. We follow the explanation before (\ref{eq:g_evolution-app}) to rewrite the evolution of the connected correlations $g_t$ as
\begin{equation}
(\partial_{t}-\Delta)g_{t}([n])=\sum_{r\sqcup s=[n]}\delta(r_{1},s_{1})\partial_{r_{1}}g_{t}(r)\partial_{s_{1}}g_{t}(s),\label{eq:g_n_evolution_compact}
\end{equation}
where $r\sqcup s=[n]$ denotes the union of non-crossing subsets $r=\{r_1,r_2,\cdots\}$ and $s=\{s_1,s_2,\cdots\}$ of the set $[n]=\{x_1,\cdots,x_n\}$. The proof of (\ref{eq:phi_equation}) is by induction over $n$.
For $n=1$, $\varphi_{t}$ and $g_{t}$ are identical and therefore
satisfy the same equation. Assume the formula holds for all $n\le k-1$
for some $k\in\mathbb{N}$. 

We start by evaluating its left hand side
making use of (\ref{eq:g_n_evolution_compact}) (for better readability
we suppress write the time argument)
\begin{align}\label{eq:lhs_terms}
(\partial_{t}-\Delta)\varphi([n])
=& \sum_{\pi\in NC([n])}\sum_{c\in\pi}(\partial_{t}-\Delta_{c})g(c)g_{\pi\backslash c}
\\ \nonumber
=& \sum_{\pi\in NC([n])}\sum_{c\in\pi}\sum_{a\cup a'=c}\delta(a_{1},a'_{1})\partial_{a_{1}}g_{t}(a)\partial_{a'_{1}}g_{t}(a')g_{\pi\backslash c}.
\end{align}
Here $\Delta_{c}=\sum_{x\in c}\Delta_{x}$ and $\pi\backslash c$
is the partition $\pi$ without the block $c$. 

We continue with the
right hand side of (\ref{eq:g_n_evolution_compact}),
\begin{align}\label{eq:rhs_terms}
 \sum_{r\sqcup s=[n]}\delta(r_{1},s_{1})\partial_{r_{1}}\varphi(r)\partial_{s_{1}}\varphi(s) 
 & =\sum_{r\sqcup s=[n]}\delta(r_{1},s_{1})\Big(\partial_{r_{1}}\sum_{\rho\in NC(r)}g_{\rho}\Big)\Big(\partial_{s_{1}}\sum_{\sigma\in NC(s)}g_{\sigma}\Big) 
 \\ 
 & =\sum_{\substack{r\sqcup s=[n]\\ \rho\in NC(r)\\ \sigma\in NC(s)}}\delta(r_{1},s_{1})\partial_{r_{1}}g(b(r_{1}))\partial_{s_{1}}g(b(s_{1}))g_{\rho\backslash b(r_{1})}g_{\sigma\backslash b(s_{1})} \nonumber
\end{align}
where the blocks $b(r_{1})\in\rho$ and $b(s_{1})\in\sigma$ are uniquely
defined by the fact that $r_{1}\in b(r_{1})$ and $s_{1}\in b(s_{1})$. 

Note that terms appearing in the sum in (\ref{eq:rhs_terms}) and
(\ref{eq:rhs_terms}) agree iff $a=b(r_{1})$, $a'=b(s_{1})$ and
$\pi\backslash c=\rho\backslash b(r_{1})\cup\sigma\backslash b(s_{1})$.
To show equivalence between the two sums, we should show that
there is a bijection between the sets 
\begin{gather*}
\{(a,a',\pi)\mid\pi\in NC([n]),a\cup a'=:c\in\pi\}\\
\longleftrightarrow\\
\{(r_{1},s_{1},\rho,\sigma)\mid\rho\in NC(r),\sigma\in NC(s),r\sqcup s=[n]\}.
\end{gather*}
\begin{enumerate}
\item[{``$\rightarrow$'':}] Given $\pi\in NC([n]),a\cup a'=:c\in\pi$, this
completely specifies a terms in the sum in (\ref{eq:g_n_evolution_compact}).
Define $r_{1}=a_{1}$ and $s_{1}=a'_{1}$. By the condition that $r\sqcup s=[n]$
is non-crossing, this uniquely defines the sets $r$ and $s$. Since
$\pi$ is non-crossing, once we take away the ``connecting block''
$c$, $\pi\backslash c$ factorizes in a unique way into two non-crossing
partitions $\tilde{\rho}$ of $r\backslash a$ and $\tilde{\sigma}$
of $s\backslash a'$. That is $\pi\backslash c=\tilde{\rho}\cup\tilde{\sigma}$.
From these we can define $\rho=\tilde{\rho}\cup\{a\}$ and $\sigma=\tilde{\sigma}\cup\{a'\}$
such that indeed, $\pi\backslash c=\rho\backslash a\cup\sigma\backslash a'$.
This produces a corresponding term in the sum of (\ref{eq:rhs_terms}),
which is completely specified by the data $(r_{1},s_{1},\rho,\sigma)$.

\item[{``$\leftarrow$'':}] Given $\rho\in NC(r),\sigma\in NC(s),r\sqcup s=[n]$,
we define $a=b(r_{1})$, $a'=b(s_{1})$ and $c=a\cup a'$ It remains
to construct $\pi\in NC([n])$ such that $\pi\backslash c=\rho\backslash a\cup\sigma\backslash a'$.
This is achieved by defining $\pi:=\rho\cup\sigma\cup\{c\}$.
\end{enumerate}

\subsection{Steady state solution}
\label{app:steady_measure}

Here, we prove that $\varphi_\infty(x_1,\cdots,x_n)=\mathrm{min}(x_{1},...,x_{n})$ is a steady state solution of \eqref{eq:phi_equation} .

We represent the minimum by a sum of Heaviside-functions $\Theta(\mathrm{condition})$
which are one if the condition is true and zero otherwise,
\[
\mathrm{min}(x_{1},...,x_{n})=\sum_{i=1}^{n}x_{i}\Theta(x_{i}<\{x_{1},\cdots,\hat{x}_{i},\cdots,x_{n}\}).
\]
The hat on $\hat{x}_{i}$ suggests that $x_{i}$ is missing from the set $\{x_{1},\cdots,x_{n}\}$. 
We have $\partial_i\mathrm{min}(x_{1},...,x_{n})=\Theta(x_{i}<\{x_{1},\cdots,\hat{x}_{i},\cdots,x_{n}\}) $.
Furthermore, derivatives of the Heaviside-function evaluate to 
\begin{align*}
\partial_{i}\Theta(x_{i}<\{x_{1},\cdots,\hat{x}_{i},\cdots,x_{n}\}) 
&= -\sum_{j\ne i}\delta(x_{i},x_{j})\Theta(x_{i}<\{x_{1},\cdots,\hat{x}_{i},\hat{x}_{j},\cdots,x_{n}\}),
\\
\partial_{j}\Theta(x_{i}<\{x_{1},\cdots,\hat{x}_{i},\cdots,x_{n}\})
&= \delta(x_{i},x_{j})\Theta(x_{j}<\{x_{1},\cdots,\hat{x}_{i},\hat{x}_{j},\cdots,x_{n}\}).
\end{align*}
With these formulas, it is easy to check that 
\begin{align*}
&-\Delta\min(x_{1},\cdots,x_{n})
=\sum_{\underset{i\neq j}{i,j}} \delta(x_{i},x_{j}) \Theta(x_{j}<\{x_{1},\cdots,\hat{x}_{i},\hat{x}_{j},\cdots,x_{n}\})).
\end{align*}
Furthermore, we have
\begin{align*}
   \delta(x_{i},x_{j}) \Theta(x_{j}<\{x_{1},\cdots,\hat{x}_{i},\hat{x}_{j},\cdots,x_{n}\})
   =\ \delta(x_{i},x_{j}) \Theta(x_{i}<\{x_{i+1},\cdots,\hat{x}_{j}\})
   \Theta(x_{j}<\{x_{j+1},\cdots,\hat{x}_{i-1}\}).
\end{align*}
As a consequence,
\begin{align*}
&-\Delta\min(x_{1},\cdots,x_{n})
=\sum_{\underset{i\neq j}{i,j}} \delta(x_{i},x_{j})\partial_{i}\min(x_{i},\cdots,x_{j-1})\partial_{j}\min(x_{j}, \cdots,x_{i-1}),
\end{align*}
This is all we needed to show the claim.
\end{widetext}


\bibliography{references}

\end{document}